\pdfoutput=1 
\documentclass[letterpaper]{elsarticle}
\usepackage{booktabs}
\usepackage{threeparttable}
\usepackage{array}
\usepackage{amsmath}
\usepackage{amssymb}
\usepackage{amsthm}
\usepackage{amsfonts}
\usepackage[caption=false,font=footnotesize]{subfig}
\usepackage{algorithm}
\usepackage{algpseudocode}
\usepackage{dblfloatfix}
\usepackage[dvipsnames]{xcolor}
\usepackage{graphicx}
\usepackage{breqn}
\usepackage{IEEEtrantools}
\usepackage{rotating}
\newcommand{\approptoinn}[2]{\mathrel{\vcenter{
			\offinterlineskip\halign{\hfil$##$\cr
				#1\propto\cr\noalign{\kern2pt}#1\sim\cr\noalign{\kern-2pt}}}}}

\definecolor{MyBlue}{rgb}{0.019,0.384,0.486}

\journal{Journal of Signal Processing}
\begin{document}
\begin{frontmatter}
	\title{Single Stage DOA-Frequency Representation of the Array Data with Source Reconstruction Capability}
	\author{Shervin~Amirsoleimani}
	\ead{amirsoleimani@ut.ac.ir}
	\author{Ali~Olfat}
	\ead{aolfat@ut.ac.ir}
	\address{Signal Processing and Communication Systems Laboratory, University of Tehran, Tehran, Iran.}
	\begin{abstract}
    In this paper, a new signal processing framework is
    proposed, in which the array time samples are represented in DOA-frequency domain through a single stage problem.
    It is shown that concatenated array data is well represented in a $\mathbf{G}$ dictionary atoms space, where $\mathbf{G}$ columns correspond to pixels in the DOA-frequency image. We present two approaches for the $\mathbf{G}$ formation and compare the benefits and disadvantages of them. A mutual coherence guaranteed $\mathbf{G}$ manipulation technique is also proposed.
    Furthermore, unlike most of the existing methods, the proposed problem is reversible into the time domain, therefore, source recovery from the resulted DOA-frequency image is possible.
    The proposed representation in DOA-frequency domain can be simply transformed into a group sparse problem, in the case of non-multitone sources in a given bandwidth. Therefore, it can also be utilized as an effective wideband DOA estimator.
    In the simulation part, two scenarios of multitone sources with unknown frequency and DOA locations and non-multitone wideband sources with assumed frequency region are examined.
    In multitone scenario, sparse solvers yield more accurate DOA-frequency representation compared to some noncoherent approaches. At the latter scenario, the proposed method with group sparse solver outperforms some existing wideband DOA estimators in low SNR regime.
    In addition, sources' recovery simultaneous with DOA estimation shows significant improvement compared to the conventional delay and sum beamformer and without prerequisites required in sophisticated wideband beamformers.
\end{abstract}

\end{frontmatter}
\date{\today}
\section{Introduction}\label{sec.int}
Wideband signal processing is an active area of research in array processing. 
It has been studied extensively in areas such as radar, seismology, sonar, radio astronomy, speech acquisition and, acoustics \cite{nordholm2014broadband,Reddy2012}.
Although, the problem of direction-of-arrival (DOA) estimation has attracted considerable attention in array processing, representation of array data in DOA-frequency domain has not been directly examined. Certainly, any narrowband or wideband DOA estimation or source localization method without knowledge of the source's frequency contents is impossible \cite{VanTrees2002}. Furthermore, in real situations, a passive wideband array receives multiple sources with different bandwidths and center frequencies. Therefore, in non-cooperative scenarios, estimating the DOA-frequency distribution of the array data seems an inevitable pre-processing.

DOA estimation of wideband sources are essentially categorized into two different approaches; \textit{noncoherent} and \textit{coherent}. In noncoherent wideband processing, each subband is processed separately and the results are combined noncoherently. While in the coherent solution, DOA estimation is performed in a single center frequency. It combines the subbands covariance matrices coherently via a \textit{focusing matrix} and the final covariance is used for DOA estimation\cite{Wang1985}. Some advantages and drawbacks of them are listed below:
\begin{itemize}
	\item
	Noncoherent approach suffers from intrinsic noncoherent losses.
	\item 
	Increasing focusing errors with the bandwidth expansion makes the performance of coherent method worse than noncoherent one in ultra-wideband scenarios.
	\item 
	The prerequisite of both solutions is the knowledge of frequency contents of the signal and dramatic performance degradation occurs in the case of incorporating noise only subbands in overall combination.
	\item 
	Coherent methods require initial focusing angles, therefore, utilizing noncoherent pre-processing or DOA-frequency estimation is vital as the first step.
\end{itemize}


Some recent methods tried to overcome these problems. R-CSM \cite{Sellone2006} proposed an iterative auto-focusing procedure to relax the requirement for initial DOA estimation. Test of orthogonality of projected subspaces (TOPS) \cite{Yoon2006a} completely removed prerequisites of focusing procedure. It tests the orthogonality of the projected signal subspace and the noise subspace at each DOA and frequency subband, and, therefore, can be categorized in the noncoherent class. TOPS shows poor performance at high SNR levels and often leads to spurious peaks at all SNRs. Other algorithms were developed to improve TOPS, such as ETOPS \cite{Zhang2010a}, Squared-TOPS \cite{Okane2010} and, WS-TOPS \cite{Hayashi2016}.

In parallel, some wideband array processing techniques exploiting the special structure of the antenna array have been developed. For example, \cite{Chan2007} addresses frequency-invariant techniques for uniform concentric circular arrays (UCCAs) with omni-directional sensors and authors in \cite{Liao2013a} extended that results for directional elements and with simpler array design procedure.

Sparse representation (SR) framework has also found new applications in DOA estimation during \cite{Yang2017}. The fundamental work of Malioutov et al \cite{malioutov2005sparse} named $\ell_1-$SVD and two recent works \cite{Hu2014,Shen2017} are categorized in sparsity-based solution for DOA estimation of the wideband signals. Although, $\ell_1-$SVD is basically a narrowband solution, but have been extended to wideband situation by exploiting joint-sparsity in frequency-DOA domain.
Authors in\cite{Hu2014} used the idea proposed in \cite{Shkolnisky2006} for approximating the band-limited signals with prolate spheroidal wave functions. It represented the array observations in a dictionary and applied block orthogonal matching pursuit (BOMP) to impose the sparsity only among bases with different delays.
Recently, two time-domain solutions \cite{liu2011direction,Hu2012a} for wideband DOA estimation employing SR have been proposed. They represent array covariance matrix in temporally delayed versions of source correlation function. This correlation function is assumed known in \cite{Hu2012a} and can be estimated from observations in \cite{liu2011direction}.

In this paper, we propose a general framework to represent the array time samples directly in the space-frequency domain. It is named direct DOA-frequency representation (DDFR).
The key difference between the proposed framework and most of the existing solutions in the wideband array processing is that unlike a two-phase approach (DOA estimation and then beamforming), DDFR solves the whole problem in a single stage, i.e., sources' frequency content, spatial locations, and recovery coefficients are obtained by solving a single under-determined linear system. In other words, it is simultaneously an estimator and also a wideband beamformer.

The significant innovation of DDFR is the introduction of a $\mathbf{G}$ dictionary, by which the array multiple snapshots can be well represented in its atoms space. In other words, each column of $\mathbf{G}$ stands for a $(\theta,f)$ pixel in two-dimensional DOA-frequency image.
Two approaches for $\mathbf{G}$ atoms arrangement are proposed and compared. It is shown that the first approach leads to a constant coherence among DOA-frequency atoms but their positions in DOA are uncontrolled. The second approach arranges atoms at desired $(\theta,f)$ points but there is no straightforward control over the dictionary coherence.
We show that in noiseless case, DOA-frequency representation is formulated as a linear system in columns space of $\mathbf{G}$ . 
Regarding the problem size, number of snapshots and required DOA-frequency resolution, this linear system would yield an under-determined system, therefore has infinite solutions. Different solvers for this linear system are examined, including conventional minimum $\ell_2$-norm constraint and also sparsity-based penalty functions $\ell_0$ and $\ell_1$.


The solution of this system contains the DOA-frequency contents of the array data. One can use elements of the solution vector to reconstruct the source at a particular region in frequency or DOA. %
It means extending super-resolution capability into spatial filtering, which is absent in conventional and modern array processing techniques. For example, in a passive sonar array, one can listen to two adjacent sources and distinguish a marine mammal from a far passing ship. Furthermore, it can be utilized to show the power density of the array signal in DOA-frequency domain and estimate sources' frequency contents for further processing.

Furthermore, in DDFR it is possible to impose sparsity only in the spatial domain, rather than in both DOA and frequency. It is necessary, especially when dealing with non-multitone wideband sources with known frequency band, such as chirp signals or band-limited Gaussian processes. In this case, the degree of freedom decreases and the problem reduces to group sparse form. Therefore, DDFR can act as a wideband DOA estimator when multiple non-multitone wideband sources with identical band exist. 
Briefly, DDFR framework does not suffer from previously mentioned noncoherent and coherent drawbacks, makes no assumption on the prior statistical information such as sources distribution or incoherence, does not involve subband processing and multiple eigenvalue decomposition (EVD), and in comparison with SR-based approaches enjoys excellent performance at low SNR regime and low number of snapshots with simultaneous beamforming feature. This special feature is obtained because the signal phase information is not lost through the estimation algorithm using DDFR.

The paper is organized as follows. Section~\ref{sec.SignalModel} reviews the narrowband and wideband array signal model. Section~\ref{sec.SSR} presents a brief introduction to the problem of sparse signal representation (SSR) and group sparse formulation. The main idea of DDFR and its related issues are presented in Section~\ref{sec.JSFD}, and in Section~\ref{sec.Simulation} the simulation results are examined for two scenarios of wideband signals. Finally, in Section~\ref{sec.Conclusion} conclusions are drawn.
\section{Signal Model}\label{sec.SignalModel}
Let $s(t)$ denote a signal impinging on a linear array with $N_S$ sensors, located at $\theta$. %
The signal is received by sensors with different delays $\tau_k$. The delay $\tau_k$ is a function of the $k$'th sensor position and the source angle of arrival $\theta$. In the far-field situation, the observation vector $\mathbf{y}(t)$ containing all sensors data samples is,
\begin{equation} \label{equ.measurement_y}
\mathbf{y}(t) = \left[\begin{array}{c}
s(t-\tau_1) \\
\vdots \\
s(t-\tau_{N_S})
\end{array}
\right] \;, \qquad
\tau_k = \frac{\mathbf{a}^T\mathbf{p}_k}{c}
\end{equation}
where $\mathbf{a}$ is a unit vector describing source's arrival direction, $\mathbf{p}_k$ is the $k$'th sensor position and $c$ is the wave propagation velocity in the medium \cite{VanTrees2002}. For a uniform linear array aligned with $z$-axis and element distance $d$, the time delay due to $k$'th sensor is $\tau_k = -(k-1)d\sin(\theta)/c$.
\vspace*{-5pt}
\subsection{Narrowband case}
In narrowband case, the source $s(t)$ can be written as $s(t) = u(t)\exp(j2\pi f_0 t)$, where $u(t)$ is the baseband signal with a bandwidth much smaller than the carrier frequency $f_0$. By this assumption, the delayed version of $s(t)$ is approximated with a phase shift.
\begin{IEEEeqnarray}{rl}\label{equ.delayToShift0}
s(t-\tau) =& u(t-\tau)\exp(j2\pi f_0(t-\tau)) \IEEEyesnumber\IEEEyessubnumber*\\
	\approx& u(t)\exp(j2\pi f_0 t)\exp(-j2\pi f_0\tau) \label{equ.delayToShift}\\
		  =& s(t)\exp(-j2\pi f_0\tau)
\end{IEEEeqnarray}
By the equivalence of time delay to phase shift in \eqref{equ.delayToShift0}, the measurement vector, in \eqref{equ.measurement_y}, for a single snapshot is reformulated as,
\begin{equation} \label{equ.observation-model}
\mathbf{y}(t) = s(t)\left[\begin{array}{c}
e^{-j2\pi f_0 \tau_1(\theta)} \\
\vdots \\
e^{-j2\pi f_0 \tau_{N_S}(\theta)}
\end{array}
\right] = 
s(t)\mathbf{v}(\theta,f_0)
\end{equation}
where $\mathbf{v}(\theta,f_0)$ is called steering vector (or array manifold). 
Denoting $k$'th entry of the array manifold by $v_k(\theta,f_0) = \exp(j\phi_k)$ for $k\in \left\lbrace 1,\cdots,N_S\right\rbrace$, $\phi_k$ is formulated as,
\begin{equation}\label{equ.steeringVector}
\setlength{\nulldelimiterspace}{0pt}
\phi_k = \left\lbrace
\begin{IEEEeqnarraybox}[\relax][c]{r's}
2\pi f_0 (p_k/c)\sin(\theta) & general linear array \\
2\pi f_0 (kd/c)\sin(\theta) & uniform linear array (ULA) \\
k\pi (f_0/f_H)\sin(\theta) & ULA with  $d=(c/2f_H)$%
\end{IEEEeqnarraybox}
\right.
\end{equation}
where $f_H$ is the observations upper frequency limit, $d$ is the elements distance, and $p_k$ is the $k$'th sensor position aligned with $z$-axis. 
If rewrite \eqref{equ.observation-model} for $K$ sources and in the presence of additive noise we will have,
\begin{IEEEeqnarray}{rCl}\label{equ.steeringMatrix}
\mathbf{y}(t) &=& \left[\mathbf{v}(\theta_1,f_0),\cdots,\mathbf{v}(\theta_K,f_0)\right] \left[\begin{array}{c}
	s_1(t) \\
	\vdots \\
	s_K(t)
	\end{array}
\right]
+ \mathbf{n}(t) \IEEEnonumber\\
&=& \mathbf{V}_{N_S \times K}(\Theta,f_0)\mathbf{s}(t)_{K\times 1} + \mathbf{n}(t)
\end{IEEEeqnarray}
where $\Theta = [\theta_1,\cdots,\theta_K]^T$ is the sources direction of arrival (DOA) vector and $\mathbf{V}(\Theta,f_0)$ is the steering matrix at $f_0$.
\subsection{Wideband Case}
In the wideband scenario, the equivalence between time delay and phase shift is no longer valid. We know that time delayed version of the signal has a Fourier transform pair as $s(t-\tau)\leftrightarrow S(f)e^{-j2\pi f\tau}$. 
In this situation, phase-shifts depends on the frequency as well as the source's DOA. 
Therefore, the observation model \eqref{equ.steeringMatrix} for the wideband case, is written at each frequency subband as,
\begin{equation}\label{equ.wideband-obs-model}
\mathbf{y}(f) = \mathbf{V}(\Theta,f)\mathbf{s}(f) + \mathbf{n}(f)\
\end{equation}
where $f\in \left\lbrace f_1,\cdots,f_{N_F}  \right\rbrace $ and $ \Theta = [\theta_1,\cdots,\theta_{N_\theta}]^T$.
\section{Signal Sparse Representation (SSR)}\label{sec.SSR}
The aim of sparse representation is to find a solution for a linear under-determined system with the minimum number of non-zero elements. Given an observation $\mathbf{y}\in \mathbb{C}^M$, and a dictionary matrix $\mathbf{D}\in\mathbb{C}^{M\times N}$ with $M < N$, we would like to find $\mathbf{x}\in \mathbb{C}^N$ that satisfies the linear system $\mathbf{y}=\mathbf{D}\mathbf{x}$ and has minimum non-zero elements. Denoting $\|\mathbf{x}\|_0$ as the number of non-zero entries of a vector $\mathbf{x}$, the sparsest solution of the above under-determined system is formulated as,
\begin{equation}\label{equ.L0-problem}
\underset{\mathbf{x}}{\min} \; \|\mathbf{x}\|_0 \quad s.t. \quad \mathbf{y}=\mathbf{D}\mathbf{x} \quad (\ell_0 \text{ problem})
\end{equation} 
The $\ell_0$ problem is combinatorial in nature, and the solution requires searching through all subsets of indices of $\mathbf{x}$. This is not tractable even for moderate values of $M$ and $N$ \cite{Donoho2006}. The most common approximation is Basis Pursuit (BP) \cite{chen2001atomic}, which replaces $\ell_0$ with its closest convex norm $\ell_1$ :
\begin{equation}\label{equ.L1-problem}
\underset{\mathbf{x}}{\min} \; \|\mathbf{x}\|_1 \quad s.t. \quad \mathbf{y}=\mathbf{D}\mathbf{x} \quad (\ell_1 \text{ problem})
\end{equation}
Extending the problem to noisy measurements case we have:
\begin{equation}\label{equ.L0-noisy}
\underset{\mathbf{x}}{\min} \; \|\mathbf{x}\|_1 \quad s.t. \quad  \|\mathbf{y} - \mathbf{D}\mathbf{x}\|_2 \le \epsilon
\end{equation}
where $\epsilon$ is an upper bound for noise, such that $\| \mathbf{n} \|_2 \le \epsilon$.
\vspace*{-5pt}
\subsection{Group Sparsity} \label{sec.GroupSparsity}
In some applications, sparsity exists among clusters of the $\mathbf{x}$ entries. It means the $\mathbf{x}$ vector is partitioned into $\mathcal{G}_N$ groups with either all zeros or all nonzeros elements, $\mathbf{x} = [\mathbf{x}_{\mathcal{G}_1}^T,\mathbf{x}_{\mathcal{G}_2}^T,\cdots,\mathbf{x}_{\mathcal{G}_N}^T]^T$. The group sparsity of $\mathbf{x}$ can be measured by an $\ell_{p,q}$ norm defined as \cite{Hu2017},
\begin{equation}
\|\mathbf{x}\|_{p,q} \triangleq \left(\sum_{i=1}^{N} \| \mathbf{x}_{\mathcal{G}_i} \|_p^q \right)^{1/q}
\end{equation}
where $p$ is the norm inside a group and $q$ is the norm among distinct groups. \eqref{equ.L0-noisy} can be expressed as a group sparse problem as,
\begin{equation}\label{equ.Lpq-problem}
\underset{\mathbf{x}}{\min} \; \|\mathbf{x}\|_{p,q} \quad s.t. \quad  \|\mathbf{y} - \mathbf{D}\mathbf{x}\|_2 \le \epsilon
\quad (\ell_{p,q} \text{ problem})
\end{equation}

\section{Direct DOA-Frequency Representation (DDFR)}\label{sec.JSFD}
 First, direct representation of the array data in a $\mathbf{G}$ dictionary column space will be formulated. This acts as a $t \mapsto [f,\theta]$ transform, which gets the array time samples and results in a 2D image in frequency-DOA through a single stage formulation. 
Meanwhile, a simple solution for construction of the $\mathbf{G}$ dictionary, named constant $\Delta\delta$ approach, is derived. Then, a closed-form expression for the $\mathbf{G}$ atoms is found. This leads to a second dictionary design approach, named direct synthesis. In the next part, the mutual coherence of the $\mathbf{G}$ dictionary is calculated and a guaranteed $\mathbf{G}$ manipulation method is proposed. Finally, source reconstruction procedure is presented.

\begin{figure}[!t]
	\centering
	\includegraphics[width=2.5in]{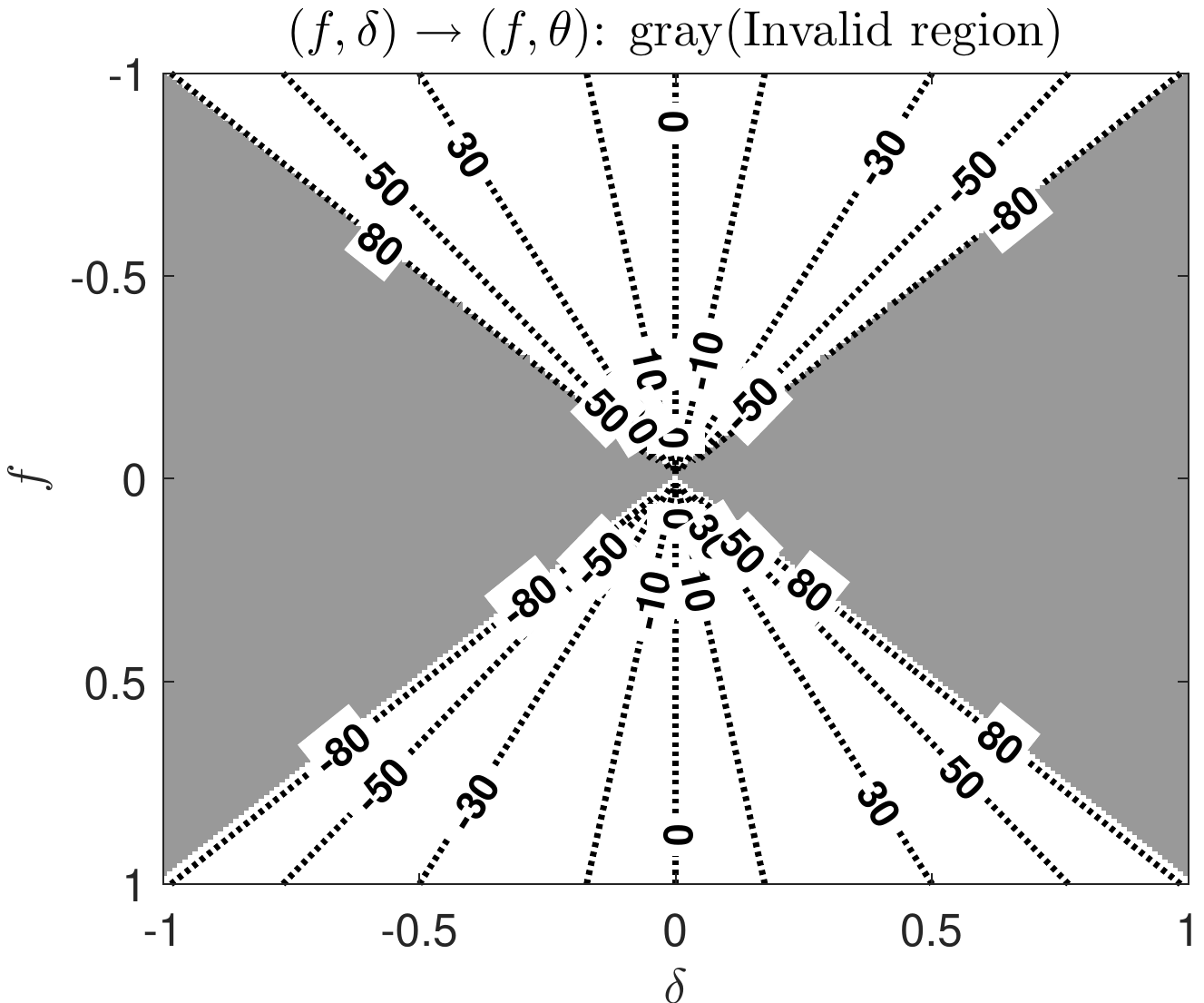}
	\caption{$(f,\delta)$ mapping to $(f,\theta)$. $f\in\left[ -1,+1\right]$,
		$\delta\in\left[ -1,+1 \right]$. The labeled dashed lines illustrate the iso-$\theta$ cells.}
	\label{Fig.f_delta_mat}
\end{figure}

Let $M$ be the number of snapshots and $\mathbf{t} = \left[t_1,\cdots , t_M \right]^T$ denote the sampling time vector. 
By arranging the array snapshots $\mathbf{y}(t)$, defined in \eqref{equ.measurement_y}, in a matrix form, the array observation matrix $\mathbf{Y}_{M\times N_S}$ is obtained as,
\begin{equation}
\mathbf{Y} = \left[\mathbf{y}(t_1),\cdots,\mathbf{y}(t_M)\right]^T
\end{equation}
where the columns of $\mathbf{Y}$ correspond to sensors' temporal samples. Denoting $\mathbf{Y}(:,n)$ as the
$n$'th column of $\mathbf{Y}$, $\mathbf{Y}(:,n)$ can be linearly represented in the Fourier bases space as follows:
\begin{equation}\label{equ.YinFourier}
\mathbf{Y}(:,n) = \sum_{k=1}^{N_F} x_k \mathbf{d}_k = \mathbf{D}(\mathcal{F})\mathbf{x}_n
\end{equation}
where $\mathbf{x}_n$ is the signal representation coefficients of the $n$'th sensors observations and
$\mathbf{D}=[d_{i,k}]$ is Fourier dictionary containing $N_F$ bases selected from $\mathcal{F}=\{f_1 , \cdots , f_{N_F}\}$ where
\begin{equation}\label{equ.fourier_atom}
d_{i,k} = \frac{1}{\sqrt{M}} \exp \left(j 2\pi f_k t_i\right)
\end{equation}
Rewriting \eqref{equ.YinFourier} in matrix form we have,
\begin{equation}\label{equ.Y=DX}
\mathbf{Y} = \mathbf{D}(\mathcal{F})\mathbf{X}
\end{equation}
where $x_{i,j}$ is the coefficient corresponding to the frequency $f_i$ in the representation of the $j$-th sensor signal. Therefore, the $k$'th row of $\mathbf{X}$ is the array snapshot in the $f_k$ frequency. Accordingly, \eqref{equ.wideband-obs-model} holds for each row of the matrix $\mathbf{X}$. By denoting $\mathbf{X}(k,:)$ as the $k$'th row of $\mathbf{X}$,
\begin{equation} \label{equ.2stage-rep}
\mathbf{X}(k,:)^T = \mathbf{V}(\Theta,f_k)\mathbf{Z}(k,:)^T \quad k\in\left\lbrace 1,\cdots , N_F \right\rbrace
\end{equation}
where $\mathbf{Z}(k,:)$ denotes the array snapshot in the $k$'th subband. 
Since the array manifold matrix $\mathbf{V}(\Theta,f_k)$ is a function of frequency as well as DOA, \eqref{equ.2stage-rep} is always examined separately in different subbands. This is the main hindrance that forces wideband array processing techniques to apply subband processing as a prerequisite. 
To resolve this problem, we rewrite steering vector as function of $\delta \triangleq f \sin(\theta)$ rather than $\theta$ or $f$.
Assume $f_L \le |f| \le f_H$ and $\theta\in[-90^\circ , +90^\circ]$, then $|\delta| \le f_H$. By this change of variable, we can rewrite steering matrix as a function of $\delta$ as,
\begin{equation}
\mathbf{V}(\Delta) = \left[\mathbf{v}(\delta_1) , \cdots , \mathbf{v}(\delta_{N_D})\right]_{N_S \times N_D}
\end{equation}
where $\Delta=\left[\delta_1,\cdots,\delta_{N_D}\right]^T$, $N_D$ is the number of $\delta$ grid points and each steering vector $\mathbf{v}(\delta_i)$ isdefined as,
\begin{equation}\label{equ.delta_atom}
\mathbf{v}(\delta_i) = \exp\left\lbrace j2\pi \delta_i \frac{\mathbf{p}}{c} \right\rbrace\:, \;\quad i\in \left\lbrace 1,\cdots,N_D\right\rbrace 
\end{equation}
For linear array, $\mathbf{p}$ is array element position vector $\mathbf{p} = \left[p_1,\cdots,p_{N_S}\right]^T$. 
Note that although $\Delta$ set leads to different DOA grid points at each $f_i$, it forms an identical steering vector matrix for the whole band. 
In other words, exploiting $\mathbf{V}(\Delta)$ in lieu of $\mathbf{V}(\Theta,f_i)$, would generate DOA grid points with identical steering matrix at all bands but with different spatial grid map. 
Rewriting \eqref{equ.2stage-rep} in matrix form and with replacing new steering matrix $\mathbf{V}(\Delta)$, we have,
\begin{equation} \label{equ.rewrite_16}
\mathbf{X}^T = \mathbf{V}(\Delta)\mathbf{Z}^T
\end{equation}
and substituting \eqref{equ.rewrite_16} in \eqref{equ.Y=DX} yields,
\begin{equation}\label{equ.Y=DZV}
\mathbf{Y} = \mathbf{D}(\mathcal{F}) \mathbf{Z} \mathbf{V}(\Delta)^T
\end{equation}
where $\mathbf{Z}_{N_F \times N_D} = [z_{i,j}]$ is the DOA-frequency representation matrix and $z_{i,j}$ corresponds to $(f_i,\delta_j)$ which is uniquely mapped to $(f_i,\theta_j)$:
\begin{equation} \label{equ.ThetaInverse}
(f_i , \delta_j ) \Rightarrow \theta_j = \sin^{-1}(\frac{\delta_j}{f_i})
\end{equation}
obviously, if $|\frac{\delta_j}{f_i}|>1$ the corresponding $z_{i,j}$ does not map to a real $\theta_j$. Let $\mathcal{S}$ be the set of all invalid $(i,j)$ pairs as,
\begin{equation}\label{equ.S_definition}
\mathcal{S} = \left\lbrace (i,j)\left| \left|\frac{\delta_j}{f_i}\right|>1  \right. \right\rbrace 
\;\Rightarrow \quad z_{i,j \in \mathcal{S}} = 0
\end{equation}
Fig.~\ref{Fig.f_delta_mat} shows the mapping from $(f,\delta)$ to $(f,\theta)$. The $\theta$ values are labeled and iso-$\theta$ contour are shown with dashed lines. Indices belonging to $\mathcal{S}$ are illustrated with gray area. With uniform $\delta$ sampling, the invalid region possesses more grid points at lower frequencies, and this will lead to a frequency dependent resolution in DOA. This issue is examined in succeeding sections.

Vectorizing \eqref{equ.Y=DZV}, can simplify the equations. Since half of the $\mathbf{Z}$ matrix's elements are zero, unacceptable entries would be removed simply in the vectorized form.
For a given matrix $\mathbf{A}\in \mathbb{C}^{m\times n}$, $\mathbf{vec}(\cdot)$ operator is defined as,
\begin{equation}\label{equ.vec-definition}
\tilde{\mathbf{a}}_{mn\times 1} \triangleq \mathbf{vec}(\mathbf{A}_{m\times n})=\left[\mathbf{a}_1^T , \cdots , \mathbf{a}_n^T\right]^T
\end{equation}
where $\mathbf{a}_k$ denotes $k$'th column of  $\mathbf{A}$. Applying  $\mathbf{vec}(\cdot)$ to \eqref{equ.Y=DZV}, we have,
\begin{equation}\label{equ.vectorize1}
\tilde{\mathbf{y}} = \mathbf{vec}(\mathbf{Y}) = \mathbf{vec}(\mathbf{D}\mathbf{Z}\mathbf{V}^T) =  
\left(\mathbf{V}\otimes \mathbf{D} \right)\mathbf{vec}(\mathbf{Z}) = 
\mathbf{G}^\prime \tilde{\mathbf{z}}
\end{equation}
where $\otimes$ stands for Kronecker Product. For $\mathbf{A}\in\mathbb{C}^{l\times n}$ and $\mathbf{B}\in\mathbb{C}^{p\times m}$ the Kronecker Product is defined as,
\begin{equation}\label{equ.kronecker}
\mathbf{B}_{p\times m}\otimes \mathbf{A}_{l\times n} \triangleq \left[\begin{array}{ccc}
b_{11}\mathbf{A}_{l\times n} & \cdots & b_{1m}\mathbf{A}_{l\times n} \\
\vdots 				& \ddots & \vdots \\
b_{p1}\mathbf{A}_{l\times n} & \cdots & b_{pm}\mathbf{A}_{l\times n}
\end{array}\right]_{pl \times mn}
\end{equation}
In \eqref{equ.vectorize1}, $\tilde{\mathbf{y}}$ is named the concatenated measurement vector, $\tilde{\mathbf{z}}$ is the resulting representation coefficients and $\mathbf{G}^\prime = \mathbf{V}\otimes \mathbf{D}$ is DOA-frequency dictionary. The $\mathbf{vec}(\cdot)$ operator should also be applied to $\mathcal{S}$ matrix as $\tilde{\mathbf{s}} = \mathbf{vec}(\mathcal{S})$. Therefore, \eqref{equ.S_definition} converts to 
$\tilde{z}_{k\in \tilde{\mathbf{s}}}=0$. This constraint is trivial since it can be applied directly to $\mathbf{G}^\prime$, by removing its invalid columns. 
Algorithm~\ref{alg.G-correction} shows a simple procedure to generate $\mathbf{G}$ from $\mathbf{G}^\prime$. In Algorithm~\ref{alg.G-correction}, $\mathbf{G}$ is the final DOA-frequency dictionary and $\tilde{\mathbf{v}}_f$ and $\tilde{\mathbf{v}}_\theta$ are frequency and DOA grid points corresponding to columns of $\mathbf{G}$ , respectively. 
Thus, with $\mathbf{G}$ dictionary, the final DOA-frequency representation can be expressed as a linear system,
\begin{equation}\label{equ.vec-JFDD-final}
\tilde{\mathbf{y}} = \mathbf{G}\tilde{\mathbf{z}}
\end{equation}

\begin{algorithm} [!t]
	\caption{Make $\mathbf{G}$ dictionary with constant $\Delta\delta$ approach}
	\label{alg.G-correction}
	\begin{algorithmic}[1]
	\Function{make\_g\_dic\_delta}{$f_L$,$f_H$,$N_F$,$N_D$}
	\State $\mathbf{v}_f \gets \left[f_1,\cdots,f_{N_F}\right]^T$
	\Comment $f_L \le f \le f_H$
	\State $\Delta \gets \left[\delta_1 , \cdots , \delta_{N_D}\right]^T$
	\Comment $|\delta| \le f_H$
	\State make $\mathbf{D}_{M\times N_F}$ dictionary
	\State make $\mathbf{V}(\Delta)_{N_S\times N_D}$
	\Comment see \eqref{equ.delta_atom}
	\State $\mathbf{G}^\prime_{M N_S \times N_F N_D} \gets \mathbf{V}\otimes \mathbf{D}$
	\State initialize $\mathcal{S}$ with $N_F \times N_D$ zero matrix
	\State initialize $\tilde{\mathbf{v}}_f$ and $\tilde{\mathbf{v}}_{\theta}$ with empty vector
	\For{$i\gets 1, N_F$}
		\For{$j\gets 1, N_D$}
			\If{$ |\frac{\delta_j}{f_i}| \le 1 $}
				\State $\mathcal{S}(i,j) \gets 1$
				\State $\tilde{\mathbf{v}}_f \gets \left[\tilde{\mathbf{v}}_f , f_i\right]$
				\State $\tilde{\mathbf{v}}_{\theta} \gets \left[\tilde{\mathbf{v}}_{\theta} , \sin^{-1}\left(\frac{\delta_j}{f_i}\right)\right]$
			\EndIf
		\EndFor
	\EndFor
	\State Initialize $\mathbf{G}$ with empty matrix
	\State $\tilde{\mathbf{s}} \gets \text{\textbf{vec}}(\mathcal{S})$
	\For{$k\gets 1, N_F N_D$}
		\If{$\tilde{s}_k = 1$}
			\State $\mathbf{G} \gets \left[\mathbf{G} \: , \: \mathbf{g}_k^\prime\right]$
			\Comment concatenate valid atoms
		\EndIf
	\EndFor
	\State \textbf{Return} $\tilde{\mathbf{v}}_f , \tilde{\mathbf{v}}_{\theta} , \mathbf{G}$
	\EndFunction
\end{algorithmic}
%
%
%
%
\end{algorithm}
 \noindent The relation \eqref{equ.vec-JFDD-final} is a fundamental result in wideband array processing. It shows that the concatenated measurement vector $\tilde{\mathbf{y}}$, lies in the columns space of the $\mathbf{G}$ dictionary. This can also be regarded as an  extension of the narrowband model for wideband systems, since different frequencies can be incorporated in the array time signal recovery without any explicit Fourier analysis. This eliminates the subband processing for wideband signals and leads to a unified approach for narrowband and wideband scenarios.
\subsection{$\mathbf{G}$ atoms formulation} \label{sec.DirectApproach}
In \eqref{equ.vec-JFDD-final}, it is shown that concatenated array time samples can be represented in the $\mathbf{G}$ columns space. In this section, we seek for the structure of $\mathbf{G}$ dictionary in more details. A closed-form relation for $\mathbf{g}(f,\theta)$ is found, consequently, in contrast to Algorithm~\ref{alg.G-correction} that there was no control on $\theta$ grid points, a direct $\mathbf{G}$ construction is presented in which arbitrary arrangement of $\mathbf{g}(f,\theta)$ atoms in $\theta$ and $f$ is provided.

To investigate the $\mathbf{G}$ matrix more closely, we refer to its definition in \eqref{equ.vectorize1} and the Kronecker product \eqref{equ.kronecker}. Assume $\mathbf{g}_q$ is the $q$'th column of $\mathbf{G}$ corresponding to $j$'th entry of $\mathcal{F}$, symbolized as $f_j$, and $r$'th element of $\Delta$, denoted as $\delta_r$. Denoting $\mathbf{G}$, $\mathbf{D}$ and $\mathbf{V}$ entries with $g,d$ and $v$ respectively, we have,
\begin{IEEEeqnarray}{rCl}
	\mathbf{G}_{MN_S\times N_F N_D} &=& \left[g_{k,q}\right] \IEEEyesnumber\IEEEyessubnumber*\\
	\mathbf{D}_{M\times N_F} &=& \left[d_{i,j}\right]	\\
	\mathbf{V}_{N_S\times N_D} &=& \left[v_{e,r}\right]%
\end{IEEEeqnarray}
Obviously,
\begin{equation}\label{equ.g2v_d}
g_{k,q} = v_{e,r} \cdot d_{i,j}
\end{equation}
Due to definition of the Kronecker product, the following relations hold for subscripts, 
\begin{IEEEeqnarray}{rCl"rCl}
	i &=& mod(k-1,M)+1 	 &  i&\in&\left\lbrace 1,\cdots,M \right\rbrace \IEEEyesnumber\IEEEyessubnumber*\\
	j &=& mod(q-1,N_F)+1 &  j&\in&\left\lbrace 1,\cdots,N_F \right\rbrace \label{equ.j-formula}\\
	e &=& \lfloor \frac{k-1}{M}\rfloor  + 1 &  e&\in&\left\lbrace 1,\cdots,N_S \right\rbrace \\
	r &=& \lfloor \frac{q-1}{N_F}\rfloor + 1 & r&\in&\left\lbrace 1,\cdots,N_D \right\rbrace \label{equ.r-formula}
\end{IEEEeqnarray}
Substituting \eqref{equ.fourier_atom} and \eqref{equ.delta_atom} in \eqref{equ.g2v_d} yields,
\begin{IEEEeqnarray}{rCl}
	g_{k,q} &=&\exp\left\lbrace j2\pi\frac{\delta_r}{c}p_e \right\rbrace \times \frac{1}{\sqrt{M}}\exp\left\lbrace j2\pi f_j t_i \right\rbrace
	\IEEEnonumber\\
	&=& \frac{1}{\sqrt{M}} \exp\left\lbrace j2\pi f_j\left( \frac{\sin(\theta_r)}{c}p_e + t_i \right) \right\rbrace
	\label{equ.dictionary_atom}
\end{IEEEeqnarray}
It shows that the $q$'th column of $\mathbf{G}$ matrix is a pixel located at $(f_j,\theta_r)$ in DOA-frequency image, where $j$ and $r$ are calculated from \eqref{equ.j-formula} and \eqref{equ.r-formula} respectively.
Now for a given $(f,\theta)$ we have the following step by step formulation for the DOA-frequency dictionary atom, denoted by $\mathbf{g}(f,\theta)$,
\begin{IEEEeqnarray}{rCl}
	\tilde{\mathbf{p}}_{MN_S\times 1} &=& \mathbf{vec}\left(\left(\mathbf{p}_{N_S\times 1}\times \mathbf{1}_{1\times M}\right)^T\right) \label{equ.ptilde}\\
	\tilde{\mathbf{t}}_{MN_S\times 1} &=& \mathbf{vec}\left( \mathbf{t}_{M\times 1} \times \mathbf{1}_{1\times N_S} \right) \label{equ.ttilde}
\end{IEEEeqnarray}
\vspace*{-10pt}
\begin{equation} \label{equ.dicAtomFormula}
\mathbf{g}(f,\theta) =
\frac{1}{\sqrt{M}}\exp\left\lbrace j2\pi f\left( \frac{\sin(\theta)}{c}\tilde{\mathbf{p}} + \tilde{\mathbf{t}} \right)  \right\rbrace
\end{equation}
where $\mathbf{p}=\left[p_1,\cdots,p_{N_S}\right]^T$ is the array elements position vector, $\mathbf{t} = \left[t_1,\cdots,t_M\right]^T$ is the time samples vector and $\mathbf{vec}(\cdot)$ operator is defined in \eqref{equ.vec-definition}.

\subsection{Dictionary Mutual Coherence}
One criterion for the uniqueness and stability of the sparsest solution in compressive sensing (CS) theory is the mutual coherence, $\mu(\mathbf{G})$. This is defined as the largest normalized inner product between dictionary columns \cite{Bruckstein2009}. For $\mathbf{G}\in \mathbb{C}^{n\times m}$,
\begin{equation}\label{equ.mutual_coherence}
\mu(\mathbf{G}) \triangleq 
\underset{1\le k,j \le m, \: k\neq j}{\max}
\frac{| \left\langle \mathbf{g}_k , \mathbf{g}_j \right\rangle |}{\|\mathbf{g}_k\|_2 \cdot \|\mathbf{g}_j\|_2}
\end{equation}
where $\langle \cdot , \cdot \rangle$ denotes the inner product.
In fact, smaller mutual coherence leads to uniqueness and stability guarantee for solutions with more non-zero entries \cite{Elad2010}. Consider two atoms from $\mathbf{G}$ at $(f_1,\theta_1)$ and $(f_2 , \theta_2)$, then substitution of \eqref{equ.dicAtomFormula} in \eqref{equ.mutual_coherence} results in a summation on real and imaginary part of the inner product as,
\begin{dmath}\label{equ.mu_sum}
\left\langle \mathbf{g}_k , \mathbf{g}_j \right\rangle = \frac{1}{N_S} \left(\Re(\mathbf{g}_1^H \mathbf{g}_2) + j\Im(\mathbf{g}_1^H \mathbf{g}_2)\right) = 
\frac{1}{MN_s}\sum_{k=1}^{MN_s} (\rho_k + j \zeta_k)
\end{dmath}
where $\rho_k$ and $\zeta_k$  can be written as,
\begin{dgroup}
\begin{dmath}
\rho_k = 
\cos\left(2\pi\frac{\Delta\delta}{c}\cdot \tilde{p}_k\right)\cos(2\pi\Delta f \tilde{t}_k) -
\sin\left(2\pi\frac{\Delta\delta}{c}\cdot \tilde{p}_k\right)\sin(2\pi\Delta f \tilde{t}_k)
\end{dmath}
\begin{dmath}
	\zeta_k = 
	\sin\left(2\pi\frac{\Delta\delta}{c}\cdot \tilde{p}_k\right)\cos(2\pi\Delta f \tilde{t}_k) -
	\cos\left(2\pi\frac{\Delta\delta}{c}\cdot \tilde{p}_k\right)\sin(2\pi\Delta f \tilde{t}_k)
\end{dmath}
\end{dgroup}
and $\Delta\delta = f_2\sin(\theta_2) - f_1\sin(\theta_1)$ , $\Delta f = f_2 - f_1$.
Utilizing the structure of $\tilde{\mathbf{p}}$ and $\tilde{\mathbf{t}}$ in \eqref{equ.ptilde} and \eqref{equ.ttilde} and by assuming uniform time steps with sampling frequency $f_s$, for a linear array the relation \eqref{equ.mu_sum} can be written as,
\begin{dgroup}\label{equ.inner_summation}
\begin{dmath}\label{equ.rho_atoms}
	\sum_{k=1}^{MN_S}\rho_k = \sum_{i=0}^{N_S-1} \cos\left(2\pi\frac{\Delta\delta}{c}\cdot p_i\right)\cdot I_M\left(2\pi\frac{\Delta f}{f_s}\right) - 
	\sin\left(2\pi\frac{\Delta\delta}{c}\cdot p_i\right)\cdot J_M\left(2\pi\frac{\Delta f}{f_s}\right)
\end{dmath}
\begin{dmath}\label{equ.zeta_atoms}
	\sum_{k=1}^{MN_S}\zeta_k = \sum_{i=0}^{N_S-1} \sin\left(2\pi\frac{\Delta\delta}{c}\cdot p_i\right)\cdot I_M\left(2\pi\frac{\Delta f}{f_s}\right) - 
	\cos\left(2\pi\frac{\Delta\delta}{c}\cdot p_i\right)\cdot J_M\left(2\pi\frac{\Delta f}{f_s}\right)
\end{dmath}
\end{dgroup}
where $I_K(x)$ and $J_K(x)$ are defined as follows,
\begin{align}
I_K(x) &\triangleq \sum_{k=0}^{K-1} \cos(kx) = \frac{\sin\left(\frac{K}{2}x\right)\cos\left(\frac{K-1}{2}x\right)}{\sin\left(\frac{x}{2}\right)}\\
J_K(x) &\triangleq \sum_{k=0}^{K-1} \sin(kx) = \frac{\sin\left(\frac{K}{2}x\right)\sin\left(\frac{K-1}{2}x\right)}{\sin\left(\frac{x}{2}\right)}
\end{align}
 In the case of a uniform linear array with element spacing $d$, coherence can be computed as,
\begin{equation} \label{equ.sum_rho_uniform}
\mu(\mathbf{G}) = \frac{1}{MN_S}
\left| \frac
	{\sin\left(\pi N_S \frac{\Delta\delta}{c}d\right)\sin\left(\pi M \frac{\Delta f}{f_s}\right)}
	{\sin\left(\pi \frac{\Delta\delta}{c}d\right)\sin\left(\pi \frac{\Delta f}{f_s}\right)}
\right|
\end{equation}
Since \eqref{equ.inner_summation} for iso-frequency atoms (i.e. $\Delta f=0$), leads to a sum on $\exp\left(\frac{2\pi}{c}\Delta\delta\cdot p_i\right)$ sequence, therefore adjacent atoms pairs having the same frequency and identical $\Delta\delta$ have similar coherence values. In this regard, $\mathbf{G}$ construction via Algorithm~\ref{alg.G-correction} automatically controls DOA grid points density to equalize mutual coherence. This will reduce density of grid points at lower frequencies and DOAs close to $\pm90^\circ$. 

Let $\mathbf{G}_\delta$ be a DOA-frequency dictionary obtained by a constant $\delta$ grid interval $\Delta\delta = c_{\delta}$. Assume $\Delta\theta(f) = \theta_2(f) - \theta_1(f)$ is the distance between two atoms pointing to $\theta_2$ and $\theta_1$ directions at frequency $f$. Using \eqref{equ.ThetaInverse} we have,
\begin{equation}
\Delta\theta(f) = \sin^{-1}\left(\frac{\delta_2}{f}\right) - \sin^{-1}\left(\frac{\delta_1}{f}\right)
\end{equation}
Dividing both sides by $\Delta\delta$ and for small values of $\Delta\delta$, we can write,
\begin{equation}
\frac{\Delta\theta(f)}{\Delta\delta} \approxeq \frac{d}{d\delta}\left( \sin^{-1}(\delta/f) \right) = \frac{1/f}{\sqrt{1-(\delta/f)^2}}
\end{equation}
Then for a constant and small $\Delta\delta$ we have the following dictionary grid resolution in DOA at frequency $f$, 
\begin{equation} \label{equ.Theta_Resolution}
\Delta\theta(f) = \left( \frac{c_\delta}{f} \right)\frac{1}{\sqrt{1 - \sin^2(\theta)}}
\end{equation}

\begin{figure*}[!t]
	\centering
	\subfloat[]{
		\includegraphics[width=0.3\linewidth]{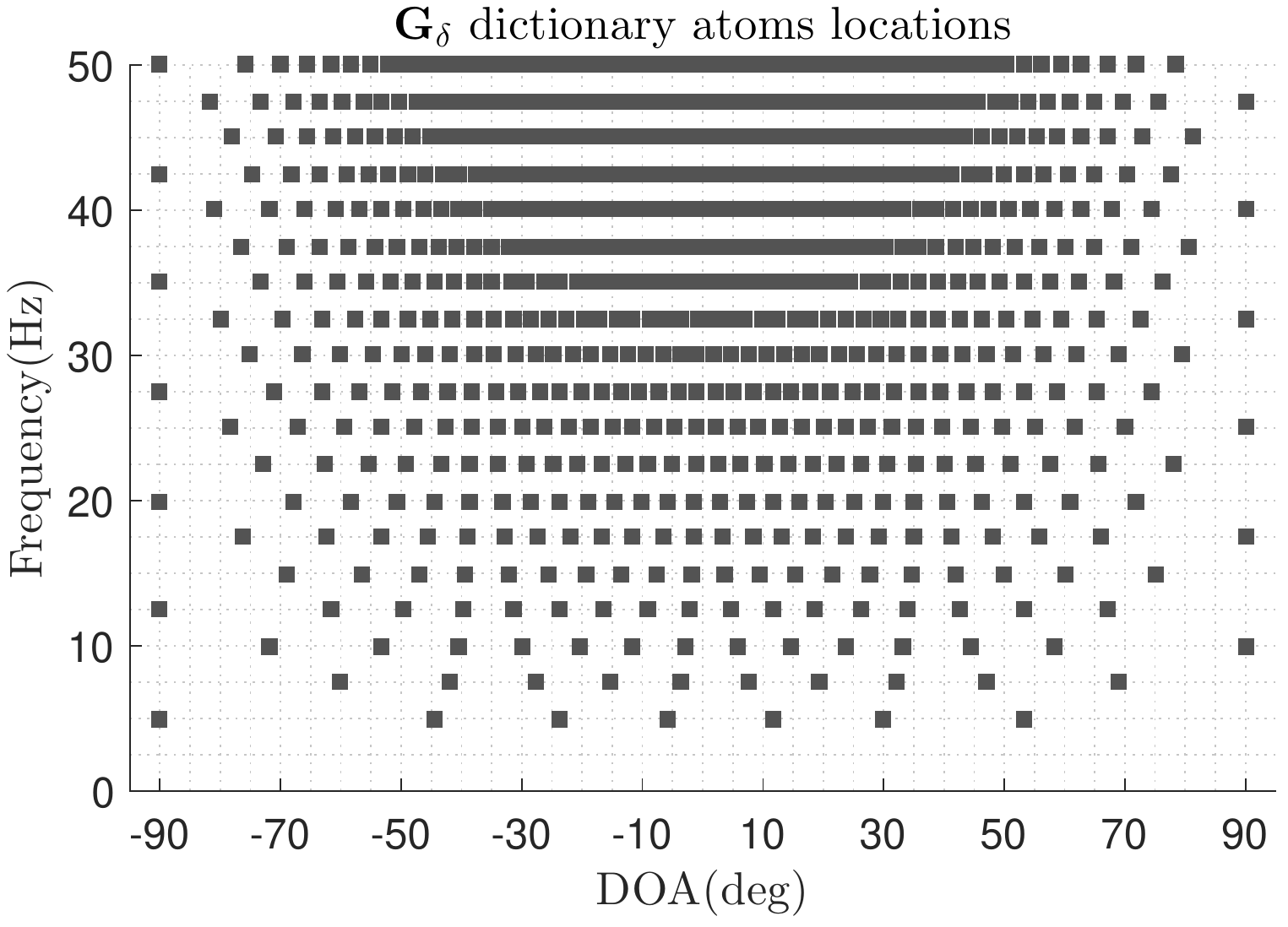}
		\label{subfig.f_delta}}
	\subfloat[]{
		\includegraphics[width=0.3\linewidth]{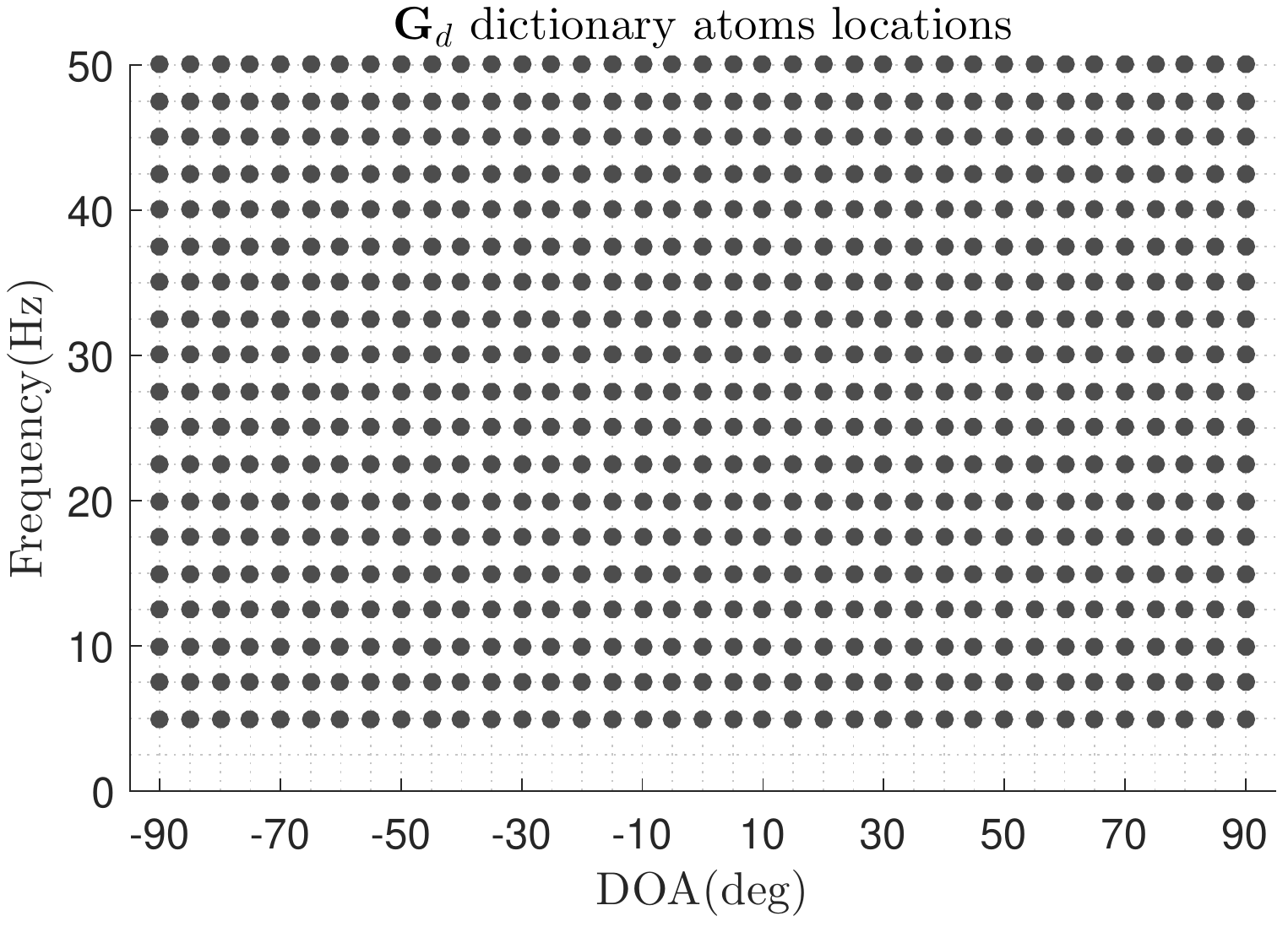}
		\label{subfig.direct}}
	\subfloat[]{
		\includegraphics[width=0.3\linewidth]{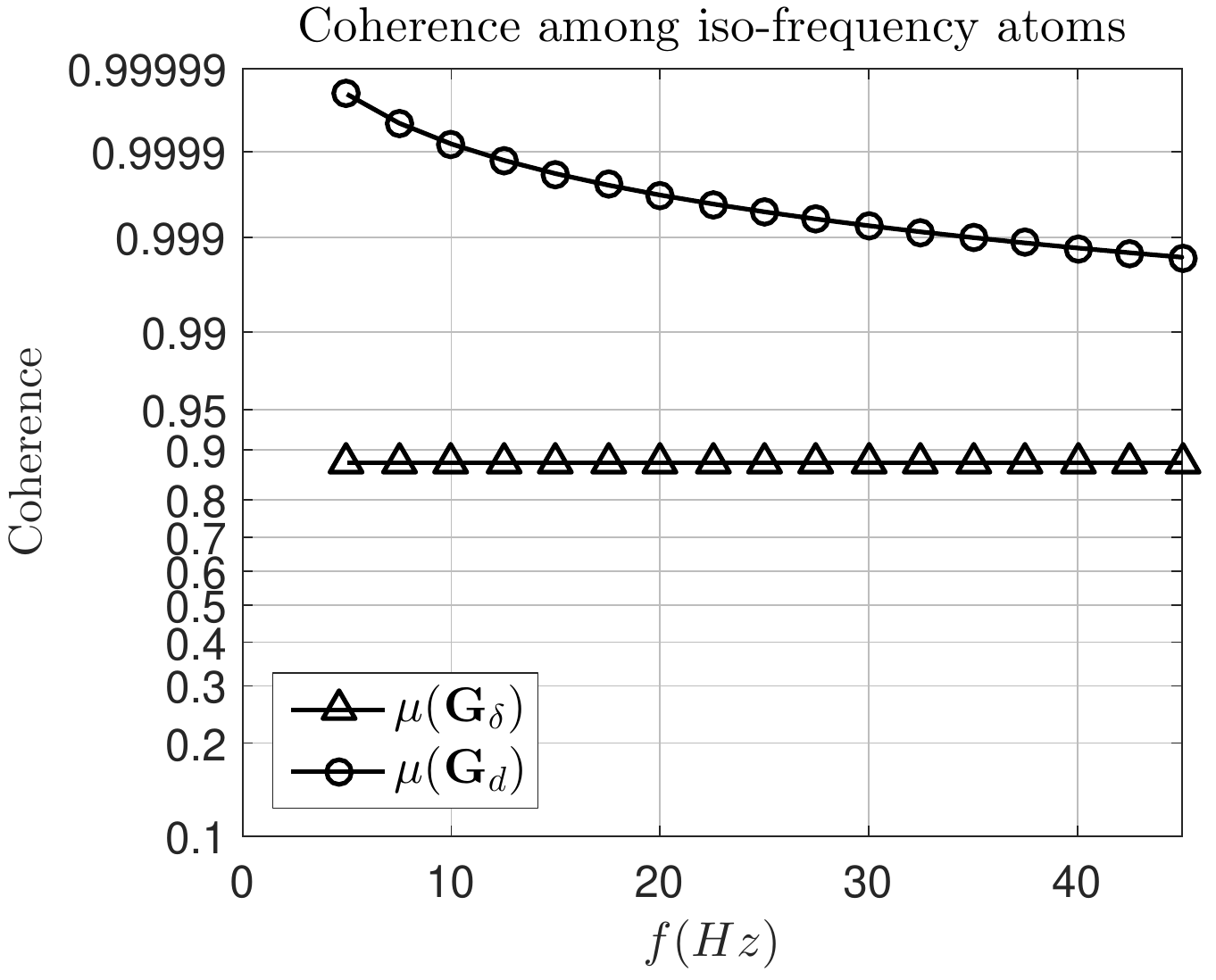}
		\label{subfig.Correlation}}
	\caption{Atoms position of $\mathbf{G}$ dictionary in DOA-frequency plane, (a) for constant $\delta$ and (b) for direct synthesis approach. The total number of atoms are the same, $f\in[5Hz , 50Hz]$ with $\Delta f=2.5Hz$. $\Delta\delta = 2.5$ for constant $\Delta\delta$ and $\Delta\theta = 8^o$ for direct approach. (c) compares coherence versus frequency for the two proposed approaches.}
	\label{Fig.G_atoms}
\end{figure*}

The relation \eqref{equ.Theta_Resolution} certifies our previous reasoning on resolution deterioration at lower frequencies and DOA borders. 
Fig.~\ref{Fig.G_atoms} shows atom's position in the $f-\theta$ plane for two dictionaries $\mathbf{G}_\delta$ (using constant $\Delta\delta$ approach, see Fig.~\ref{subfig.f_delta}) and $\mathbf{G}_d$ (using direct synthesis approach, see Fig.~\ref{subfig.direct}) with identical atoms number. In Fig.~\ref{subfig.f_delta} for the constant $\Delta\delta$ approach, density of points decreases at borders of $\theta$ to guarantee constant atoms coherence as mentioned before. On the other hand constant $\Delta\delta$ approach, density of points increases at higher frequencies near the array boresight in comparison with direct synthesis.
In Fig.~\ref{subfig.Correlation} maximum coherence among iso-frequency atoms versus frequency is shown. As expected, we see a frequency independent coherence for $\mathbf{G}_\delta$ versus a frequency decreasing correlation for $\mathbf{G}_d$. 
\vspace*{-5pt}
\subsection{$\mathbf{G}_\delta$ construction with guaranteed coherence}
As mentioned in the previous section, $\mathbf{G}_\delta$ enjoys the interesting property of constant mutual coherence among iso-frequency atoms. But $\mu(\mathbf{G})$ refers to maximum coherence among all existing atoms. In this section we propose a simple procedure to design a $\mathbf{G}_\delta$ dictionary with guaranteed coherence $\mu_0$ over all existing atoms. We present the algorithm for ULA, but it is simply extendable for any general linear array. 

With a given ULA and a desired coherence $\mu_0$, $\Delta\delta$ is obtained through solving iso-frequency atoms coherence equation \eqref{equ.sum_rho_uniform} at $\Delta f=0$:
\begin{equation}\label{equ.ddelta_solve}
\mu(\mathbf{G})\left|_{\Delta f =0}\right. = \mu_0 = \frac{1}{N_S}
\left| \frac
{\sin\left(\pi N_S \frac{\Delta\delta^*}{c}d\right)}
{\sin\left(\pi \frac{\Delta\delta^*}{c}d\right)}
\right|
\end{equation}
and for setting $\Delta f$, coherence between iso-$\delta$ atoms is considered, where $\Delta\delta=0$ is replaced in \eqref{equ.sum_rho_uniform},
\begin{dmath}
\left. \mu(\mathbf{G})\left|_{\Delta\delta =0}\right.=\mu_0\right.
=\frac{1}{M}
\left| \frac
{\sin\left(\pi M \frac{\Delta f^*}{f_s}\right)}
{\sin\left(\pi \frac{\Delta f^*}{f_s}\right)}
\right|
\end{dmath}
Finally, Algorithm~\ref{alg.G-correction} constructs a dictionary with uniform $\delta$- spacing $\Delta\delta^*$ and uniform frequency interval $\Delta f^*$.

\subsection{Source Reconstruction} \label{sec.SourceRecon}
A noteworthy property of the direct DOA-frequency representation is that it preserves the time domain data and therefore, source recovery is possible. On the contrary, in the regular array processing techniques based on signal and noise subspace properties \cite{schmidt1986MUSIC} or singular value decomposition (SVD) \cite{malioutov2005sparse}, the source time domain information is lost and super-resolution capability in DOA estimation can not be extended to source extraction. 

For a general complex measurement $\tilde{\mathbf{y}}\in \mathbb{C}^{MN_S}$ for linear system \eqref{equ.vec-JFDD-final}, Algorithm~\ref{alg.SourceExt} explains how to recover signal for a given rectangular region in $\theta-f$. In Algorithm~\ref{alg.SourceExt}, $\tilde{\mathbf{v}}_f$ and $\tilde{\mathbf{v}}_{\theta}$ denote frequencies and DOAs corresponding to $\mathbf{G}$ columns, resulted from Algorithm~\ref{alg.G-correction}. $\tilde{\mathbf{z}}$ is the solution of the linear system \eqref{equ.vec-JFDD-final}. $\mathbb{T}_R$ and $\mathbb{F}_R$ are the desired intervals in DOA and frequency respectively. $\eta$ is a threshold determining the minimum amplitude of atoms incorporated in reconstruction process. $f_s^{new}$ is an arbitrary recovery sampling frequency and $t_1$ and $t_M$ are the starting and ending times.
\begin{algorithm}[!t] 
	\caption{Source extraction using DDFR results in a rectangular region.}
	\label{alg.SourceExt}
	\begin{algorithmic}
\Function{src\_ext\_rect}{$\tilde{\mathbf{v}}_f$,$\tilde{\mathbf{v}}_{\theta}$,$\tilde{\mathbf{z}}$,$\mathbb{T}_R$,$\mathbb{F}_R$,$\eta$,$f_s^{new}$,$t_1$,$t_M$}
\Statex
Indices in  rectangle region $f\in\mathbb{F}$ and $\theta\in\mathbb{T}$ and with amplitude greater than arbitrary threshold $\eta$ are counted.
\State $\mathbb{K} \gets \left\lbrace k\mid\: |\tilde{z}|_k\ge\eta ,\: \tilde{v}_{f,k}\in\mathbb{F} ,\:
\tilde{v}_{\theta,k}\in\mathbb{T} \right\rbrace$
\Statex
\Comment generate a new time vector with $f_s^{new}$
\State $\mathbf{t}^\prime \gets \left[t_1 , t_1 + \tfrac{1}{f_s^{new}} , t_1 + \tfrac{2}{f_s^{new}} , \cdots , t_M\right]^T$
\State $\hat{\mathbf{s}} \gets \mathbf{0}$
\ForAll{$k\in\mathbb{K}$}
\State $\hat{\mathbf{s}} = \hat{\mathbf{s}} + \frac{\tilde{z}_k}{\sqrt{M}}\exp\left\lbrace j2\pi\tilde{v}_{f,k} \mathbf{t}^\prime \right\rbrace$
\EndFor
\If{real data} \Comment Consider negative frequencies
\State $\hat{\mathbf{s}} = 2Re\left\lbrace \hat{\mathbf{s}} \right\rbrace$
\EndIf
\State \textbf{Return} $\hat{\mathbf{s}}$
\EndFunction
\end{algorithmic}
\end{algorithm}

\subsection{Wideband DOA estimation with DDFR} \label{sec.WDOA_estimation}
\begin{figure}
	\centering
	\includegraphics[width=3.0in]{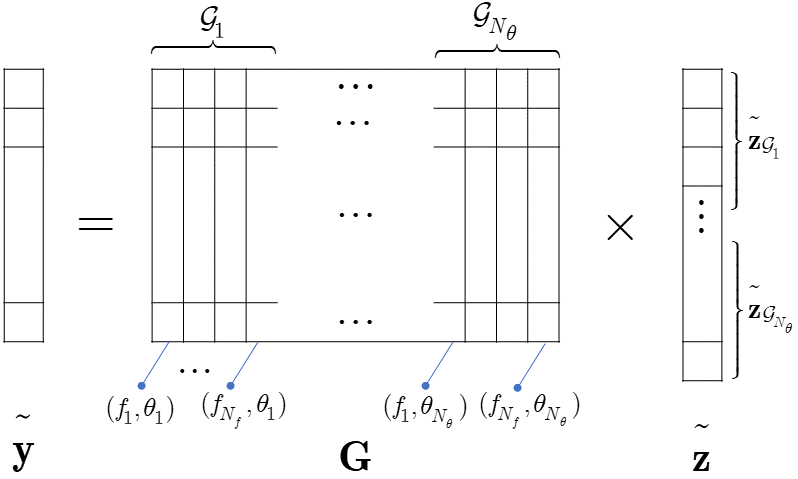}
	\caption{Using DDFR in the group sparsity formation with partitioning $\mathbf{G}$ columns and $\tilde{\mathbf{z}}$ entries into $\mathcal{G}_{N_\theta}$ groups.}
	\label{Fig.GroupForm}
\end{figure}

To employ the DDFR framework in DOA estimation problems, a naive solution is to non-coherently average $\tilde{z}_k$ at each DOA to obtain a spatial spectrum. But as discussed in Section~\ref{sec.GroupSparsity}, in the case of non-multitone sources, the group sparsity formulation can be applied to the linear system \eqref{equ.vec-JFDD-final}. Since the sources are not sparse in frequency domain but are located at sparse DOA ‌angles, the $\mathbf{G}$ dictionary atoms are partitioned into $N_\theta$ groups, in such a way that atoms with identical DOA lie in the same group. In this regard, $\mathbf{G}$ dictionary construction with constant $\Delta\delta$ method (explained in Algorithm~\ref{alg.G-correction}) cannot be used, because it does not control the position of DOA grid points. Therefore, $\ell_{p,q}-$problem \eqref{equ.Lpq-problem} can be applied and spatial spectrum at $\theta_i$ is equal to $\|\tilde{\mathbf{z}}_{\mathcal{G}_i}\|_2$. Fig.~\ref{Fig.GroupForm} illustrates this partitioning.

About the computational complexity of the algorithm it should be noted that, on the one hand, it depends on the numerical solver selection, and on the other hand, it increases with the size of the problem. The linear system $\tilde{\mathbf{y}}=\mathbf{G}\tilde{\mathbf{z}}$ dimensions increase linearly with the number of snapshots, i.e. $\mathcal{O}(M)$, it has also a linear relation with the number of grid points in DOA or frequency and also the number of sensors. In the next Section, a numerical example is presented to investigate the computational cost of the proposed method versus the number of snapshots.
\section{Numerical Simulation}\label{sec.Simulation}

In this section, numerical examples are presented to examine the performance of the proposed method. Two scenarios for the observations are considered. In the first scenario, there are multiple multitone sources impinging on the array from unknown directions and with unknown frequency contents. In this scenario, the sparsity assumption simultaneously holds at DOA and frequency domain and either of $\ell_0-$problem \eqref{equ.L0-problem} and $\ell_1-$problem \eqref{equ.L1-problem} can be applied. 
In the second scenario, there are multiple wideband Gaussian distributed sources at unknown directions but with known and identical frequency contents. This assumption is almost always applied in wideband DOA estimation methods. In this situation, the group sparsity approach is used, since the sources are not sparse at frequency domain but are located at sparse DOA ‌angles. Therefore, $\ell_{p,q}-$problem \eqref{equ.Lpq-problem} can be exploited with indexing the atoms at identical DOA in the same group. In this regard, $\mathbf{G}$ dictionary manipulation with constant $\Delta\delta$ method (explained in Algorithm~\ref{alg.G-correction}) cannot be used in the second scenario, since it does not control the position of DOA grid points.
As before, $\mathbf{G}$ dictionary with constant $\Delta\delta$ approach is denoted by $\mathbf{G}_\delta$ and $\mathbf{G}_d$ stands for $\mathbf{G}$ dictionary constructed with direct synthesis approach \eqref{equ.dicAtomFormula}.
\vspace*{-5pt}
\subsection{Wideband multitone sources}

As mentioned above, any sparse solver can be applied in this scenario. We use CVX \cite{cvx} and YALL1 \cite{YALL1_basic} solvers for the $\ell_1-$problem, Orthogonal Matching Pursuit (OMP) \cite{Rezaiifar} for $\ell_0-$problem and truncated singular value decomposition (TSVD) \cite{Hansen1990} and simple correlator estimator as a solution to $\ell_2$-problem objective function. Although $\ell_2$-problem approaches are not categorized as sparse solutions, but are illustrated for comparison. For a linear system $\mathbf{y}=\mathbf{D}_{m\times n}\mathbf{x}$ with $\text{rank}(\mathbf{D})<min\lbrace m,n\rbrace$, TSVD removes the near-zero singular values of $\mathbf{D}$ and the corresponding eigenvectors, then finds the solution by pseudo-inverse, $\mathbf{x}_{TSVD} = \mathbf{D}_{TSVD}^\dag \mathbf{y}$, where $\mathbf{D}=\mathbf{US}\mathbf{V}^H$ is the singular value decomposition of $\mathbf{D}$ and pseudo-inverse is computed as,
\begin{equation}
\mathbf{D}_{TSVD}^\dag = \mathbf{V}(:,1:T)\text{diag}([1/\sigma_1,\cdots,1/\sigma_T]) \mathbf{U}(:,1:T)^H
\end{equation}
In the correlator estimator, $\mathbf{x}^*=\mathbf{D}^H\mathbf{y}$ is given as the solution of the linear system.

As pointed out in Section~\ref{sec.int}, most of the existing wideband DOA estimation methods cannot estimate the DOA-frequency distribution of the array data. In this regard, three non-coherent methods are also simulated; Incoherent MUSIC (IMUSIC) \cite{Kailatha1984}, Incoherent Capon (ICapon) and conventional beamforming (CBF).
These methods utilize the conventional subband processing shown in Fig.~\ref{Fig.f_delta_mat}. For CBF the spatial spectrum at $f_j$ sub-band is calculated as follows,
\begin{equation}
P_{CBF}(f_j , \theta) = 
\frac{1}{W}\sum_{k=1}^{W} \left| \mathbf{v}^H(\theta,f_j) \mathbf{y}^{\mathcal{F}}_{k,j} \right|^2
\end{equation}
where $\mathbf{v}(\theta,f_j)$ stands for steering vector at $f_j$ and $\theta$ (see \eqref{equ.steeringVector}), and $\mathbf{y}^{\mathcal{F}}_{k,j}$ is the discrete Fourier transform of measurement in $k$'th section and $j$'th frequency subband, and the total samples are divided into $W$ sections.
\vspace*{-5pt}
\begin{equation}
\mathbf{y}^{\mathcal{F}}_{k,j} = \left[ y^{\mathcal{F}}_{k,j}(1),\cdots , y^{\mathcal{F}}_{k,j}(N_S) \right]^T
\end{equation}
For IMUSIC the spatial spectrum at $f_j$ is computed as,
\begin{equation}
P_{IMUSIC}(f_j , \theta) = 
\frac{1}{\left| \mathbf{v}^H(\theta,f_j)  \hat{\mathbf{U}}_n(f_j) \hat{\mathbf{U}}_n^H(f_j) \mathbf{v}(\theta,f_j) \right| }
\end{equation}
and for ICapon:
\begin{equation}
P_{ICapon}(f_j , \theta) = 
\frac{1}{
\mathbf{v}^H(\theta,f_j) 
\hat{\mathbf{S}}_{\mathbf{y}^{\mathcal{F}}\mathbf{y}^{\mathcal{F}}}^{-1}(f_j)	
\mathbf{v}(\theta,f_j)}
\end{equation}
where $\mathbf{U}_n(f_j)$ is the estimated noise subspace at $f_j$ and $\hat{\mathbf{S}}_{\mathbf{y}^{\mathcal{F}}\mathbf{y}^{\mathcal{F}}}(f_j)$ is the estimated covariance matrix at $f_j$. Subband covariance matrix and the corresponding noise subspace is calculated using eigenvalue decomposition (EVD) as follows,
\vspace*{-5pt}
\begin{align}
\hat{\mathbf{S}}_{\mathbf{y}^{\mathcal{F}}\mathbf{y}^{\mathcal{F}}}(f_j) &= 
\frac{1}{W}\sum_{k=1}^{W} \mathbf{y}^{\mathcal{F}}_{k,j} {\mathbf{y}^{\mathcal{F}}_{k,j}}^H = 
\hat{\mathbf{U}}(f_j)\hat{\mathbf{D}}(f_j)\hat{\mathbf{U}}^H(f_j)\\
\hat{\mathbf{U}}(f_j) &=\left[\hat{\mathbf{U}}_s(f_j) , \hat{\mathbf{U}}_n(f_j)\right]\\
\hat{\mathbf{D}}(f_j) &= \text{\textbf{diag}}([\hat{\sigma}^1_s,\cdots,\hat{\sigma}^{p}_s,\hat{\sigma}^1_n,\cdots,\hat{\sigma}^{N_S - p}_n])
\end{align}
$\sigma_s^k$ and $\sigma_n^k$ stand for corresponding signal and noise eigenvalues respectively.
For IMUSIC, the number of sources at each subband is assumed known. Of course, this assumption is not practical in real applications, but order estimation at each subband with minimum description length (MDL) \cite{Rissanen1978} criterion leads to disappointing results for IMUSIC.
\begin{table*}[!t]
	\renewcommand{\arraystretch}{1.3}
	\caption{Properties of wideband multitone signals.}
	\label{table.multitone_signals}
	\centering
	\begin{tabular}{|c|c|c|c|c|c|c|c|}
	\hline  & DOA & SNR(dB) & $f_1(Hz)$ & $f_2(Hz)$ & $f_3(Hz)$ & $f_4(Hz)$ & $f_5(Hz)$ \\ 
	\hline $s_1(t)$ & $-20^o$ & 10 & 15 & 16 & 20 & 30 & 31 \\ 
	\hline $s_2(t)$ & $-0^o$  & 10 & 20 & 21 & 22 & 30 & 31 \\ 
	\hline $s_3(t)$ & $+30^o$ & 10 & 40 & 41 & 42 & 43 & 44 \\ 
	\hline $s_4(t)$ & $+50^o$ & 10 & 10 & 20 & 30 & 40 & 50 \\ 
	\hline 
\end{tabular} 	
\end{table*}

\begin{figure*}[!t]
	\centering
	\subfloat[norm-$\ell_1$(CVX), $\mathbf{G}_\delta$ dictionary]%
	{\includegraphics[width=0.33\linewidth]{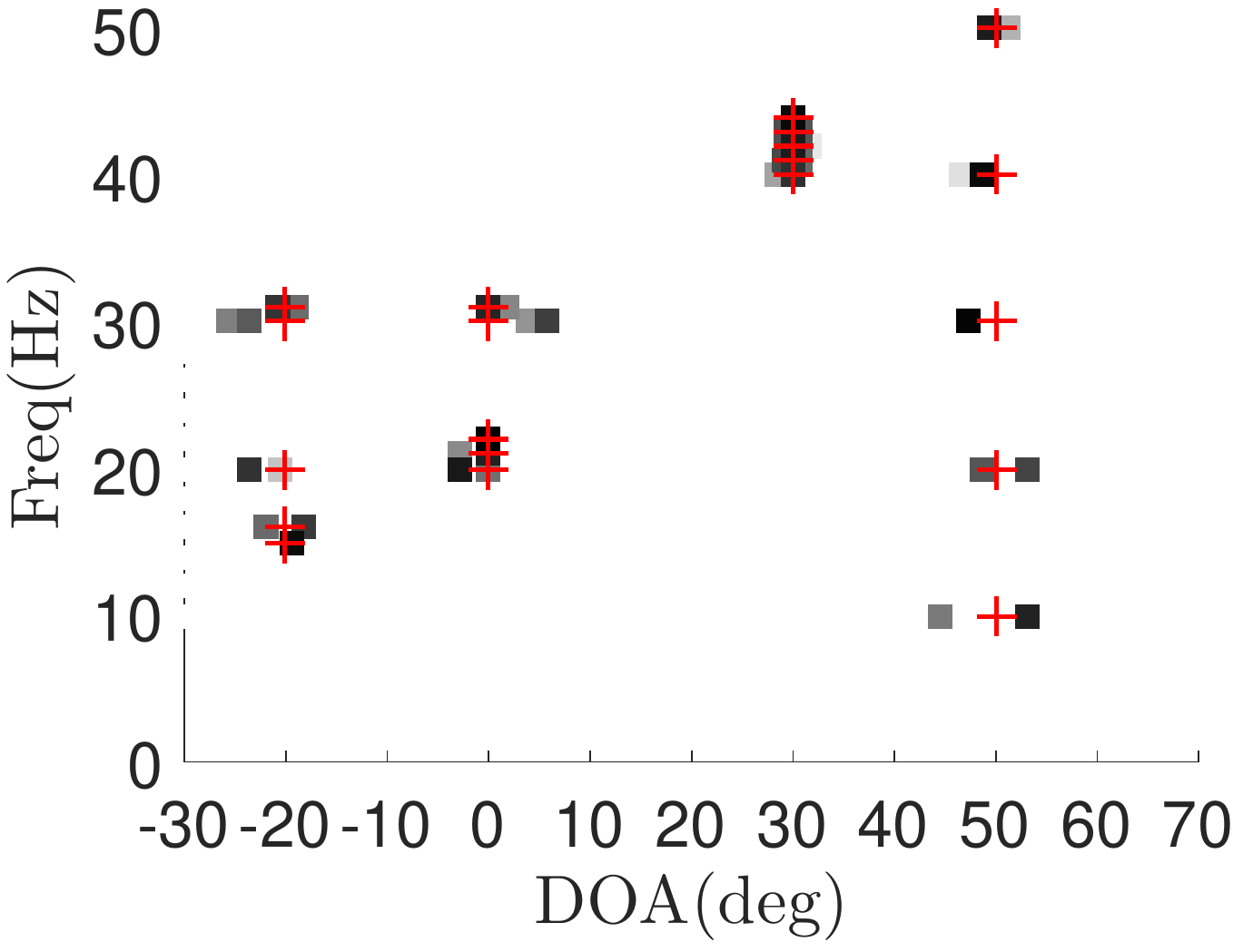}}
	\hfil
	\subfloat[norm-$\ell_1$(CVX), $\mathbf{G}_d$ dictionary]%
	{\includegraphics[width=0.33\linewidth]{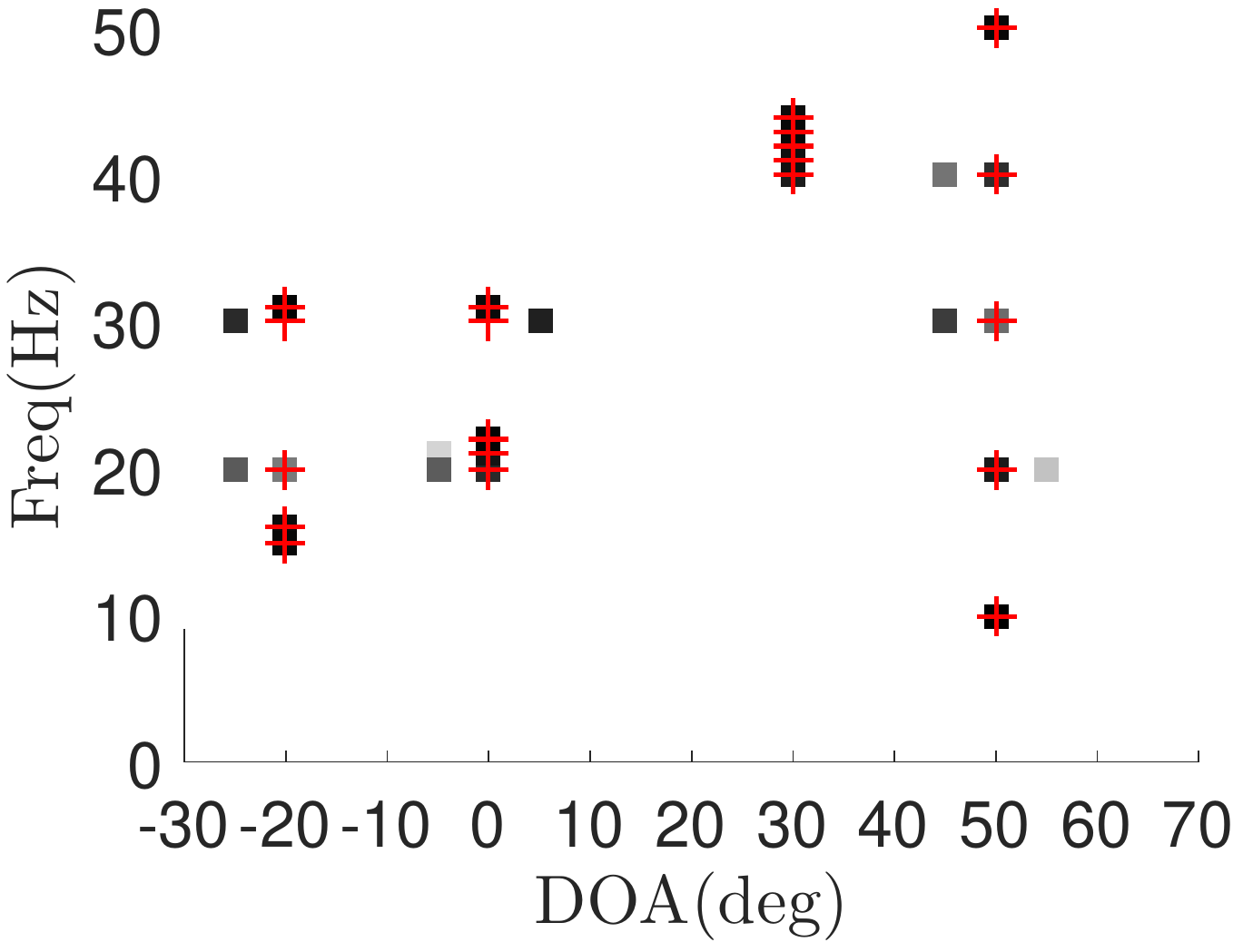}}
	\hfil
	\subfloat[norm-$\ell_1$(YALL1), $\mathbf{G}_\delta$ dictionary]%
	{\includegraphics[width=0.33\linewidth]{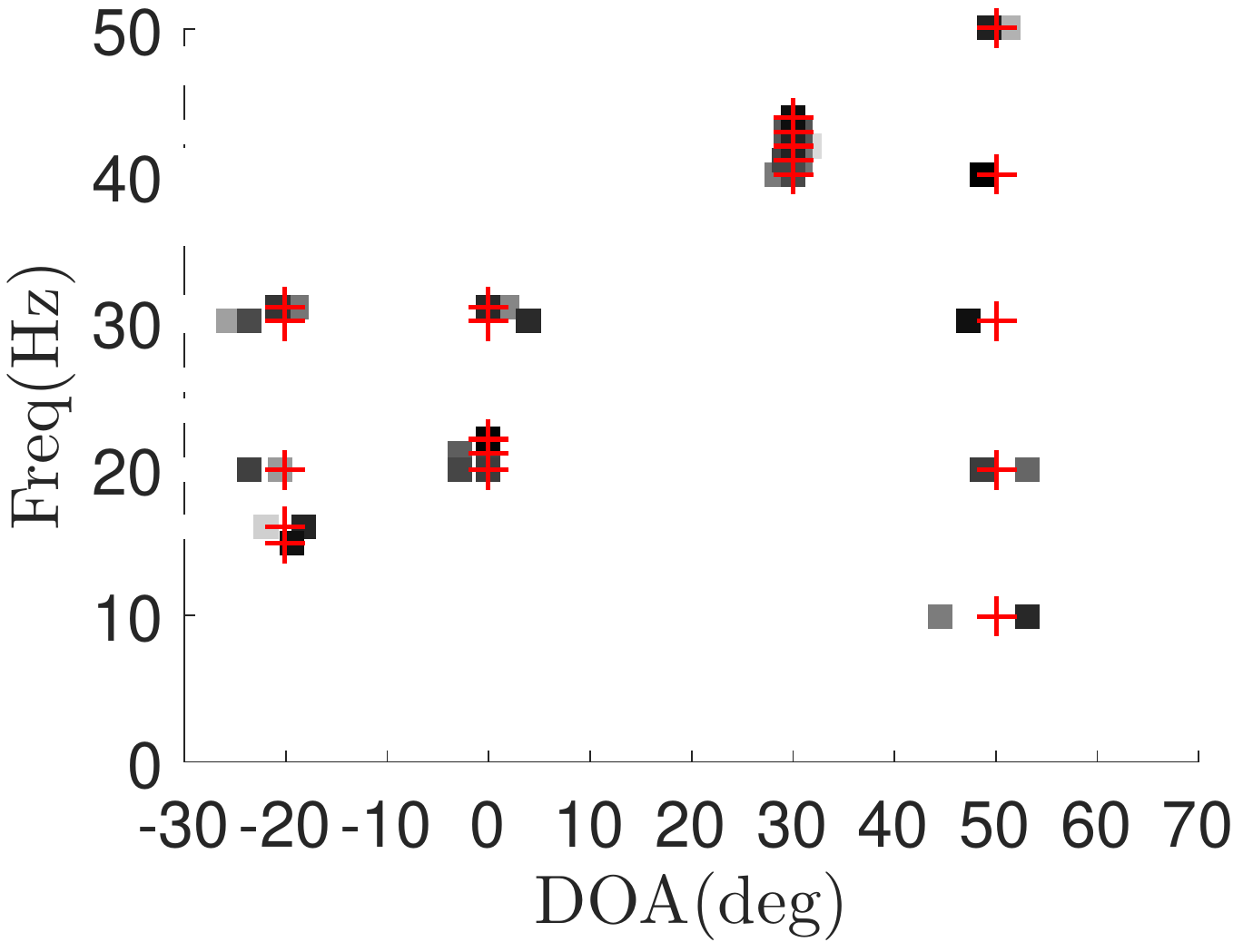}}
	\hfil
	\subfloat[norm-$\ell_1$(YALL1), $\mathbf{G}_d$ dictionary]%
	{\includegraphics[width=0.33\linewidth]{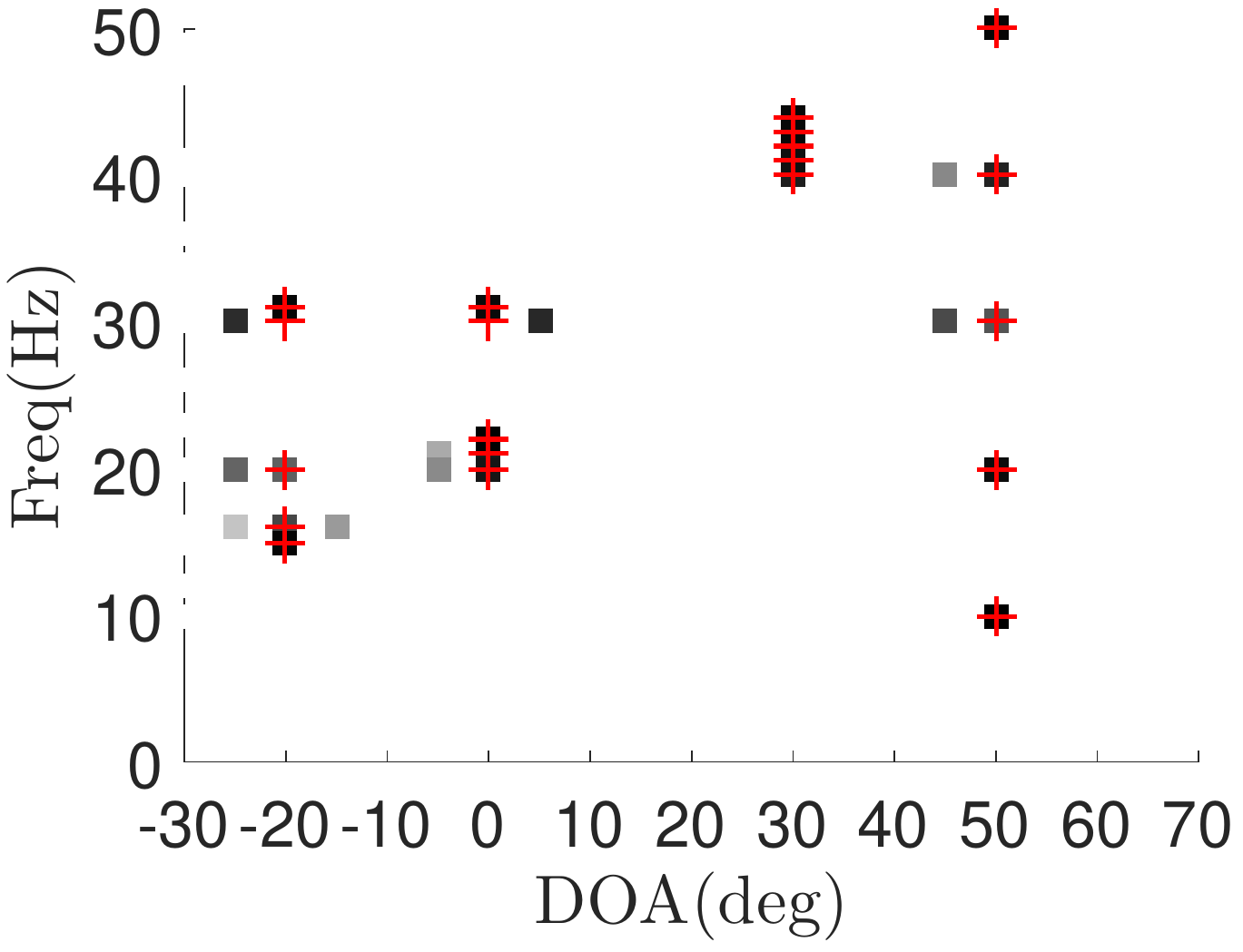}}
	\hfil
	\subfloat[norm-$\ell_0$(OMP), $\mathbf{G}_\delta$ dictionary]
	{\includegraphics[width=0.33\linewidth]{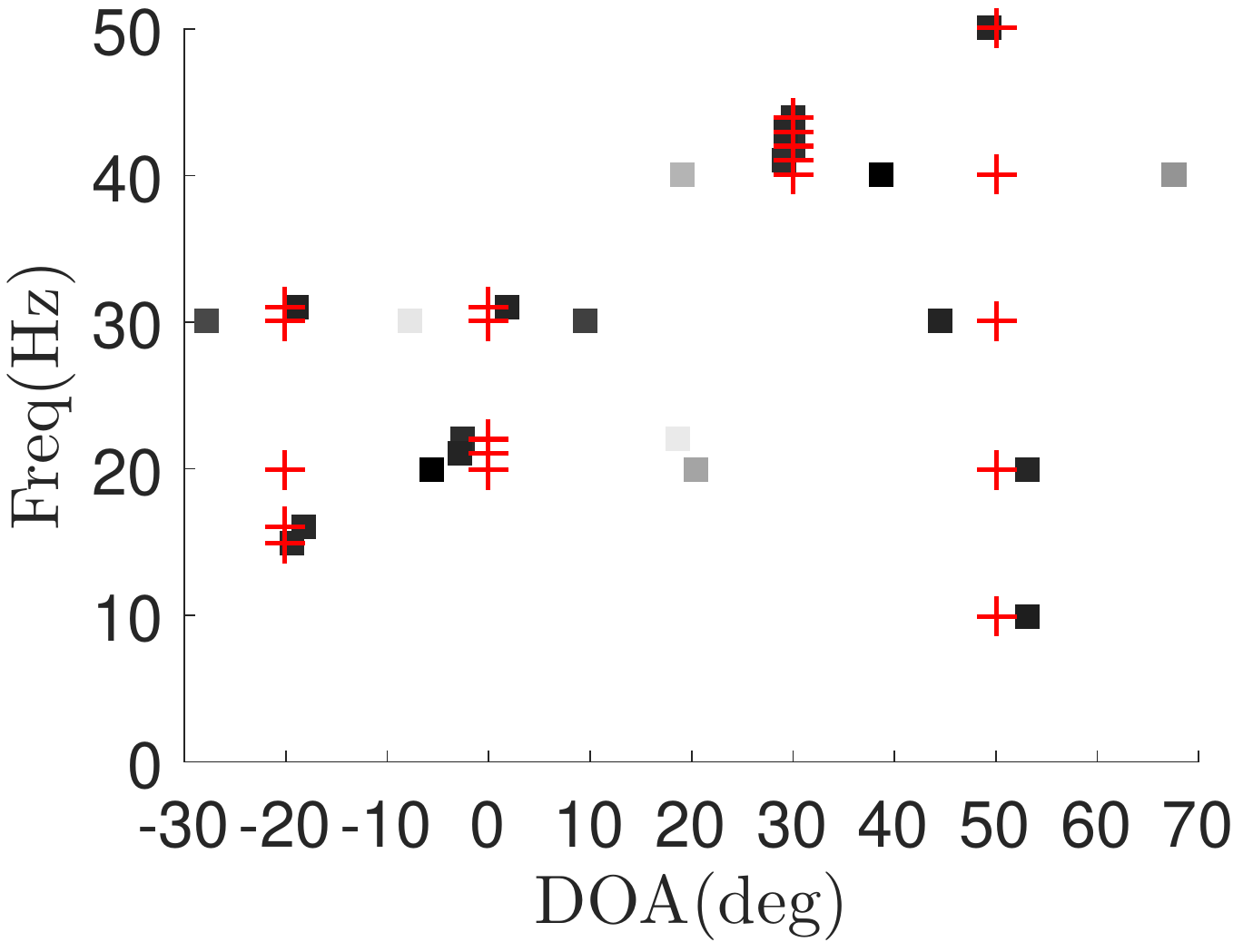}}
	\hfil
	\subfloat[norm-$\ell_0$(OMP), $\mathbf{G}_d$ dictionary]
	{\includegraphics[width=0.33\linewidth]{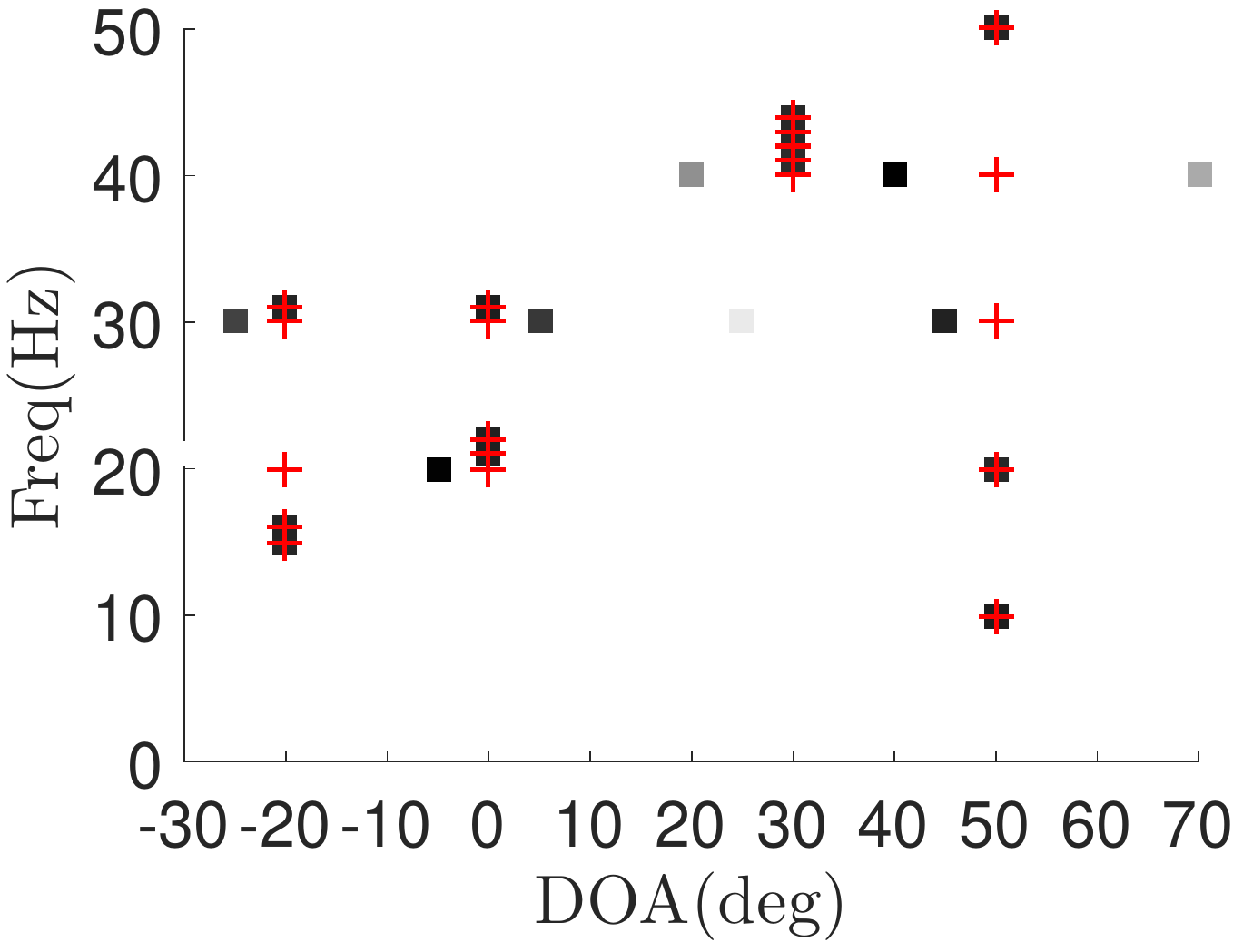}}
	\hfil
	\subfloat[norm-$\ell_2$(TSVD), $\mathbf{G}_d$ dictionary]
	{\includegraphics[width=0.33\linewidth]{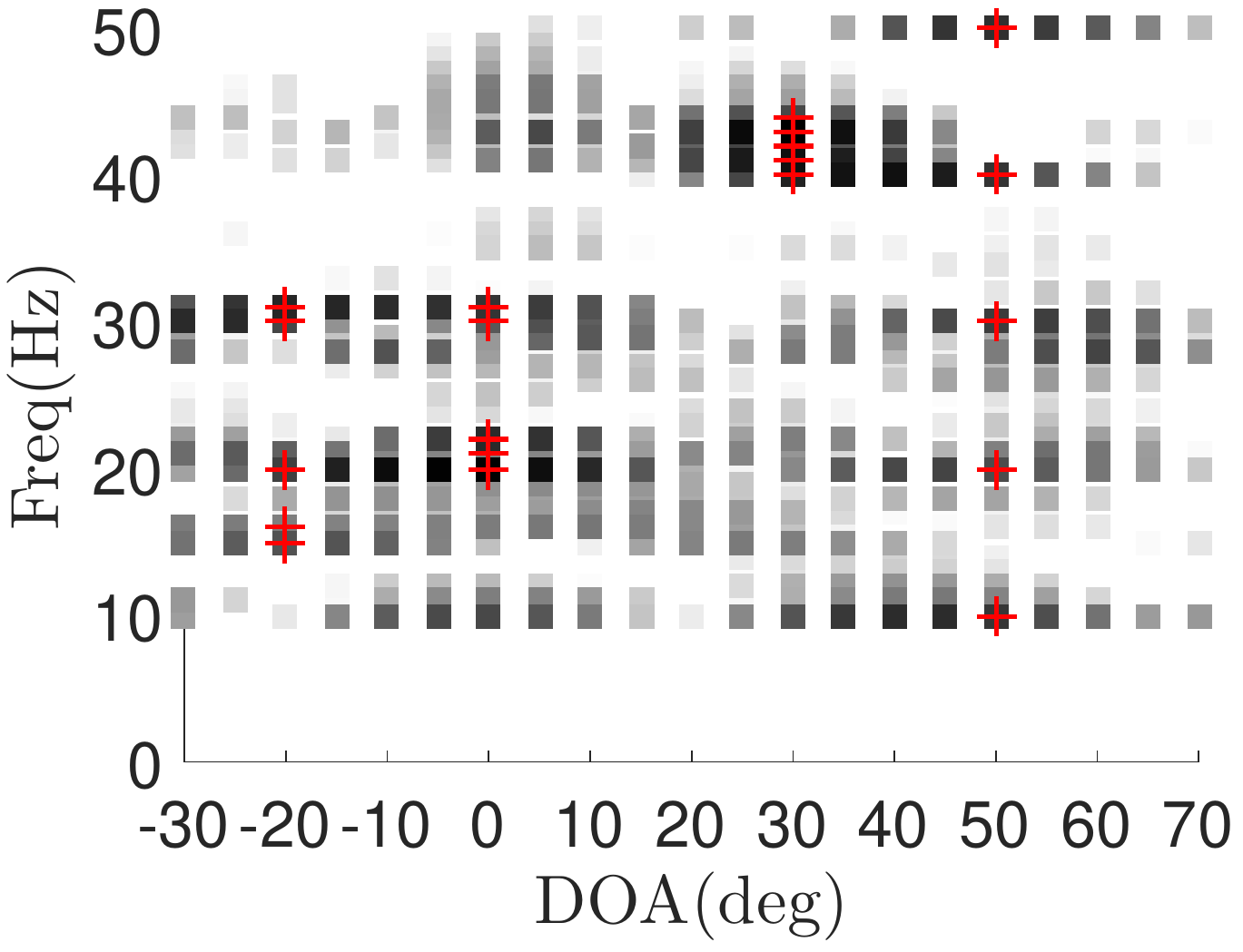}}
	\hfil
	\subfloat[norm-$\ell_2$(Correlation), $\mathbf{G}_d$ dictionary]
	{\includegraphics[width=0.33\linewidth]{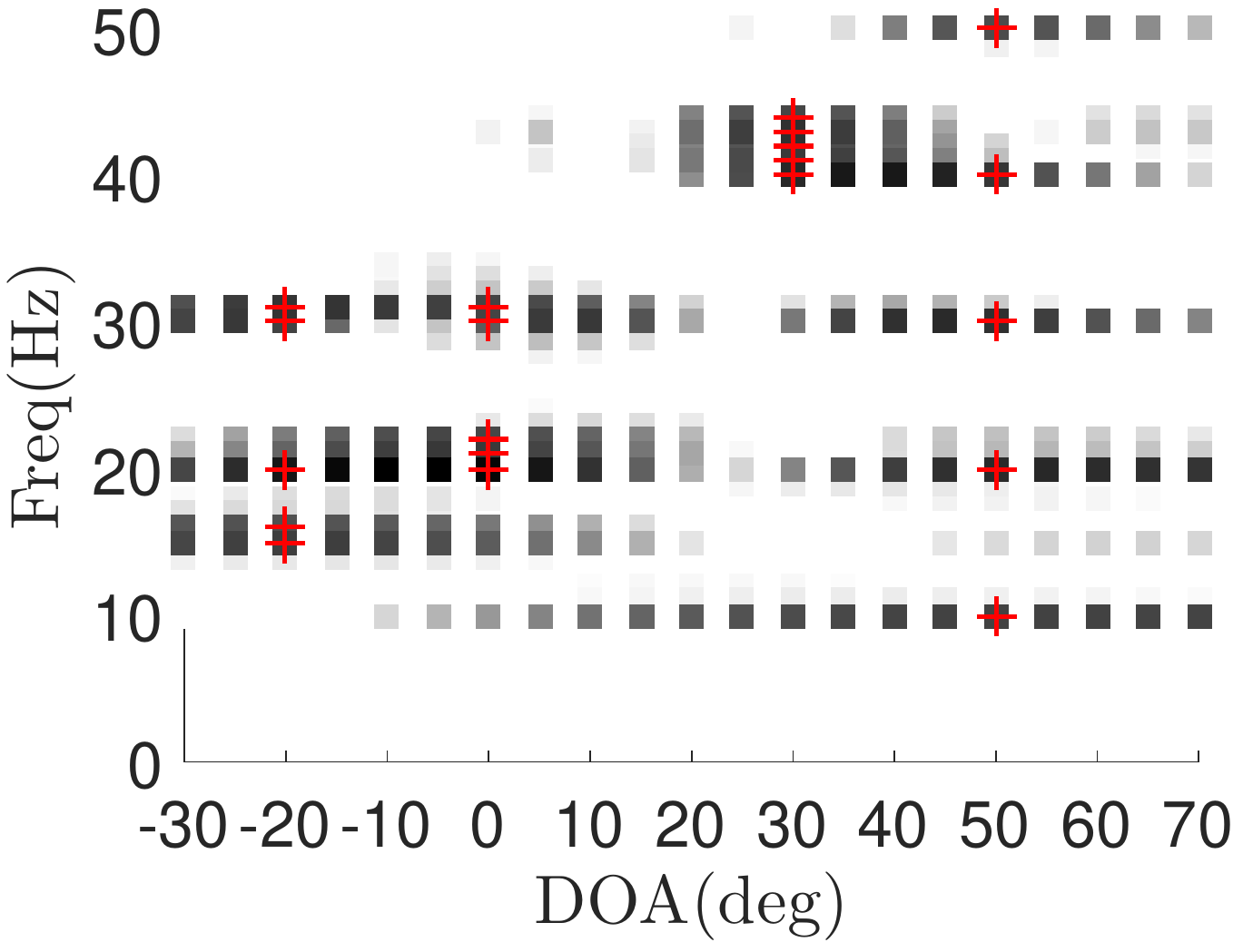}}
	\hfil
	\subfloat[Conventional beamforming]
	{\includegraphics[width=0.33\linewidth]{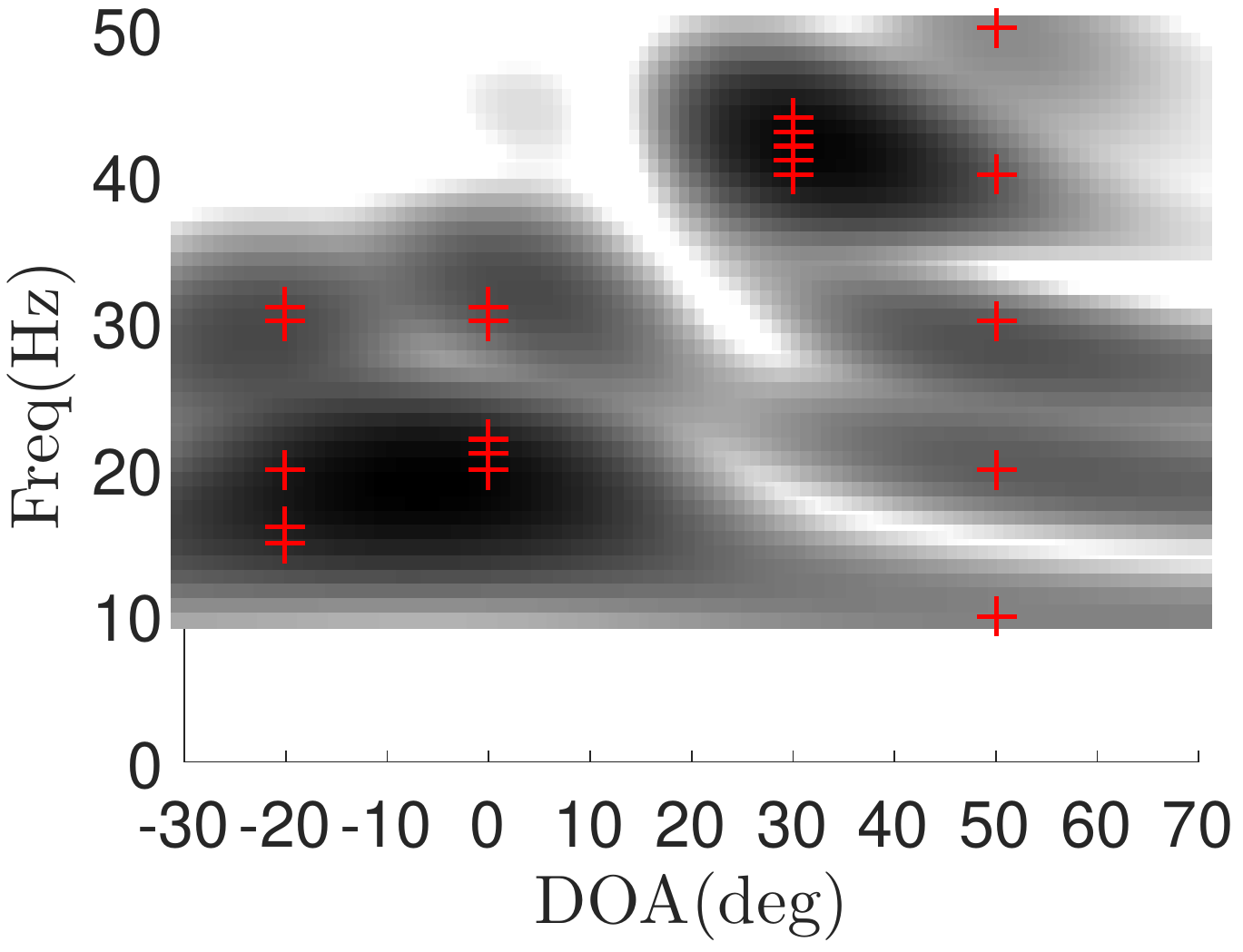}}
	\hfil
	\subfloat[ICapon]
	{\includegraphics[width=0.33\linewidth]{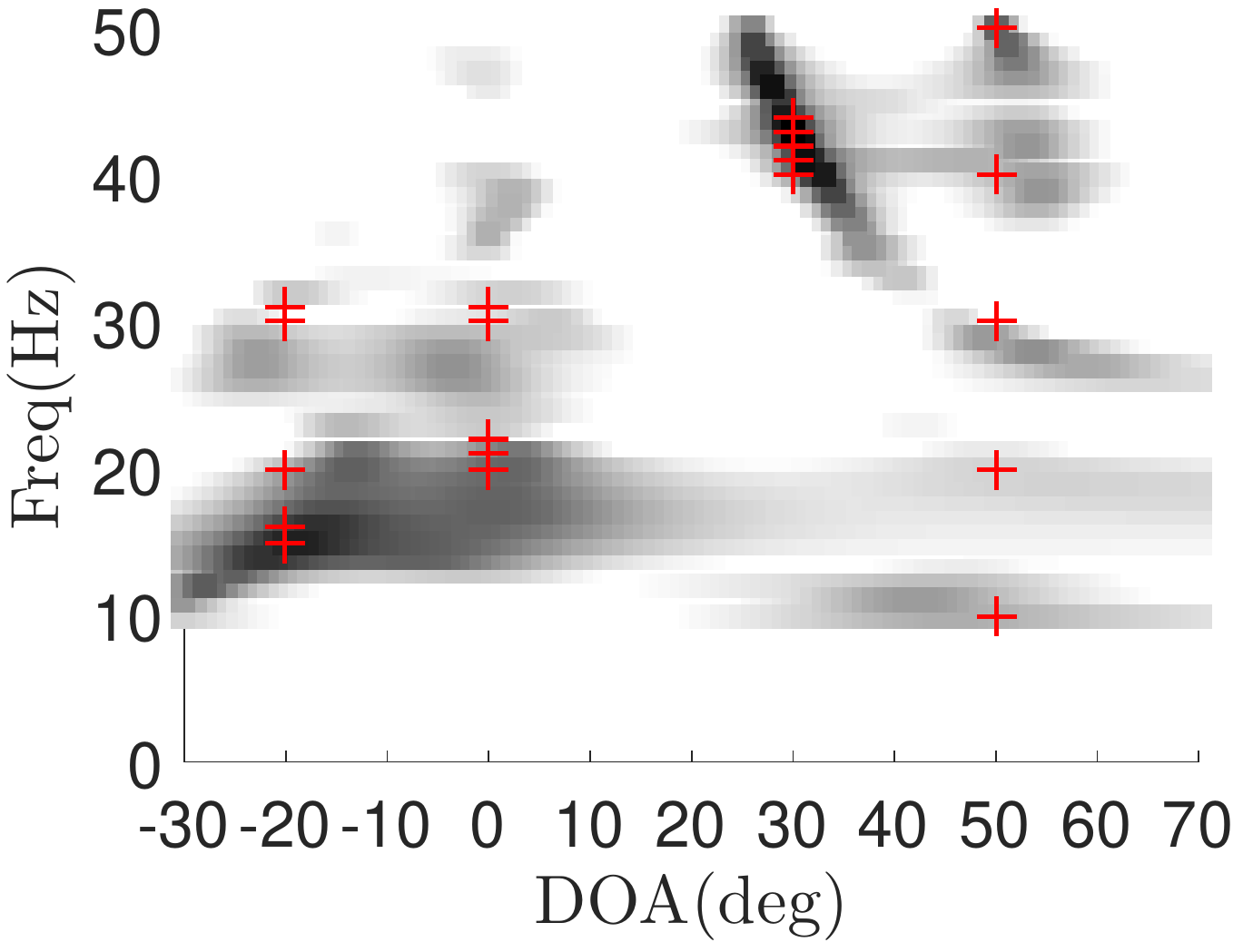}}
	\hfil
	\subfloat[IMUSIC, with known source order at each frequency sub-band]
	{\includegraphics[width=0.33\linewidth]{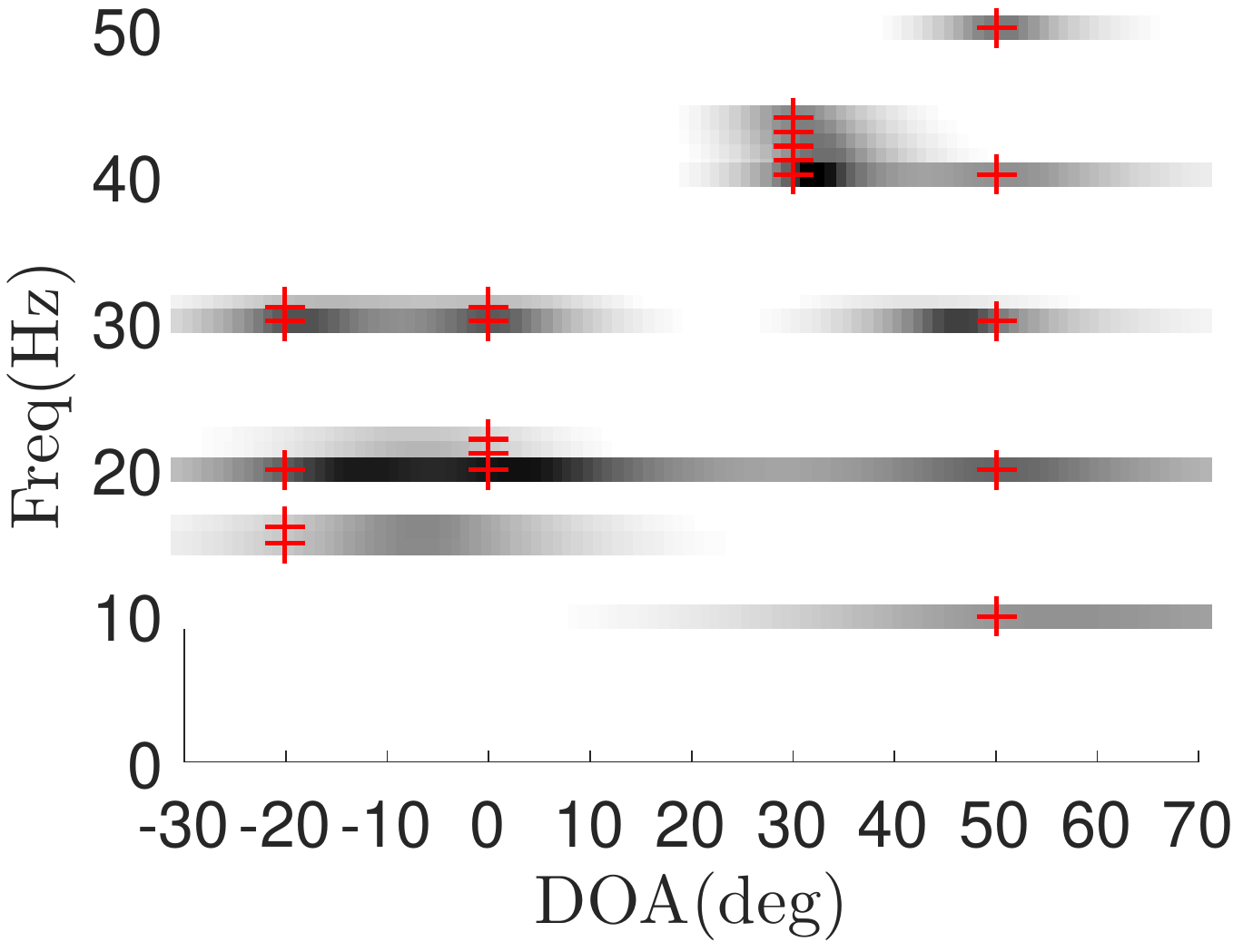}}
	\hfil
	\subfloat[IMUSIC, with unknown source order]
	{\includegraphics[width=0.33\linewidth]{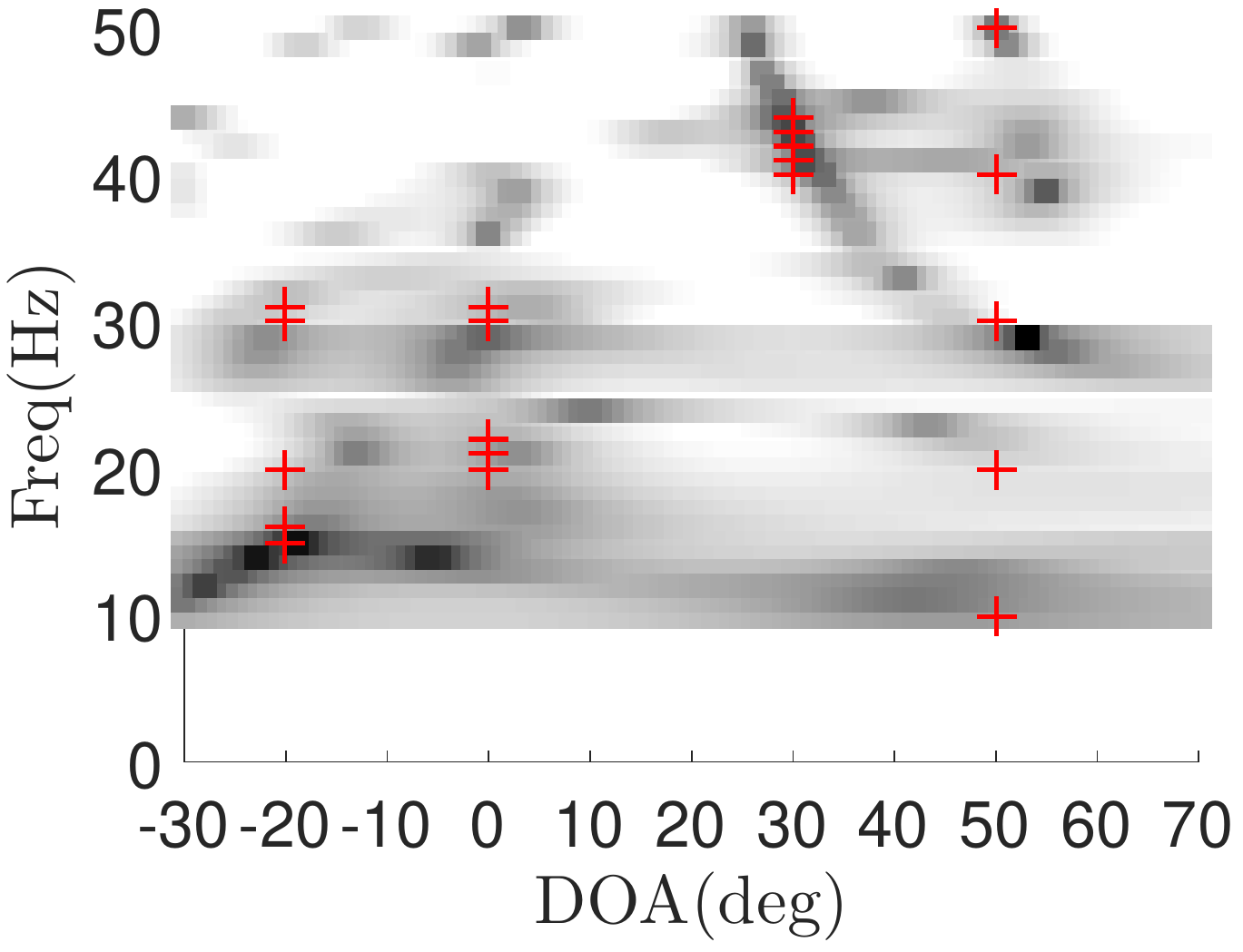}}
	\caption{Comparison of different approach to the space-frequency analysis of 4 Multi-tone wideband signals(see Table~\ref{table.multitone_signals} for signal's details). Simulation is for a ULA with 8 sensors and 100 number of snapshots. True sources' tones are plotted with red $+$ markers.}
	\label{Fig.fig1: SFD images}
\end{figure*}

Assume a ULA with 8 elements and spacing $d=\lambda_{min}/2$, where $\lambda_{min} = c/f_H$ and $c=1500m/s$ which is the underwater sound propagation velocity. The received signal spectral content is in $f\in \left[f_L , f_H\right]$, where $f_L$=10Hz and $f_H$=50Hz.
Four multi-tone sources are assumed and each source contains 5 tones. There are $M=100$ snapshots with sampling frequency $f_s=120Hz$. Sources properties are summarized in Table~\ref{table.multitone_signals}. 
The resulting DOA-frequency images are illustrated in Fig.~\ref{Fig.fig1: SFD images}. It shows that norm-$\ell_1$ approach has more accurate results in comparison with the other ones. Norm-$\ell_2$ based results show a blurred spot near real tones but suffer from low resolution and high sidelobe level. The norm-$\ell_0$ approach, which utilizes OMP solver, is orders of magnitude faster than norm-$\ell_1$ but has wrong peaks. 
As mentioned in section~\ref{sec.SourceRecon}, DDFR extends super-resolution into spatial filtering. To measure the performance of source recovery, the output signal to interference and noise ratio (SINR) relative to the input SINR is calculated. Since this improvement is obtained through an array processing technique, it can be referred to as \textit{array gain},
\begin{equation}\label{equ.array_gain}
\text{Array Gain} \triangleq \frac{\text{SINR}_{out}}{\text{SINR}_{in}} = 
\frac{\| y(t) - s(t) \|_2^2}{\| \hat{s}(t) - s(t) \|_2^2}
\end{equation}
Array gain for all solvers and the conventional Delay\&Sum beamformer is shown in Table~\ref{table.result_multitone_signals}.
It is assumed that Delay\&Sum beamformer is steered to true source locations at all subbands. 
For each source, array gain larger than that of delay\&sum beamformer is shown in bold characters. 
It is clear that $\ell_2$-norm is not a proper approach for this problem. OMP with $\mathbf{G}_d$ dictionary shows improvement on average compared to delay\&sum beamformer, whereas OMP with $\mathbf{G}_\delta$ shows poorer performance compared to delay\&sum. The best performance belongs to $\ell_1$-norm, where $\mathbf{G}_d$ shows a slight improvement over $\mathbf{G}_\delta$ and also YALL1 seems better than CVX.
\begin{table*}[!t]
	\caption{Extracted source array gain in {\upshape dB}, for different methods and multitone wideband signals of Table~\ref{table.multitone_signals}.}
	\label{table.result_multitone_signals}
	\centering
	\begin{tabular}{|c|c|c|c|c|c|c|c|c|c|}
	\hline  & \begin{sideways} {\tiny Delay\&Sum} \end{sideways}%
			& \begin{sideways}{\tiny $\ell_1$(CVX)-$\mathbf{G}_\delta$}\end{sideways}
			& \begin{sideways}{\tiny $\ell_1$(CVX)-$\mathbf{G}_d$}\end{sideways}
			& \begin{sideways}{\tiny $\ell_1$(YALL1)-$\mathbf{G}_\delta$}\end{sideways}
			& \begin{sideways}{\tiny $\ell_1$(YALL1)-$\mathbf{G}_d$}\end{sideways}
			& \begin{sideways}{\tiny $\ell_2$(TSVD)-$\mathbf{G}_d$}\end{sideways}
			& \begin{sideways}{\tiny $\ell_2$(Crr)-$\mathbf{G}_d$}\end{sideways}
			& \begin{sideways}{\tiny $\ell_0$(OMP)-$\mathbf{G}_\delta$}\end{sideways}
			& \begin{sideways}{\tiny $\ell_0$(OMP)-$\mathbf{G}_d$}\end{sideways} \\ 
	\hline $s_1(t)$ & 10.9& \textbf{15.4} & \textbf{17.1} & \textbf{16.4} &\textbf{18}  & 0.7  & -4.5 & 9.2   &  \textbf{11.4} \\
	 
	\hline $s_2(t)$ & 9.6 & \textbf{14.2} & \textbf{16.8}&\textbf{17.4}&\textbf{17.4}& -2.1 & -4.9 & 9.2   &  \textbf{15.4} \\ 
	
	\hline $s_3(t)$ & 11.1& \textbf{28.3} & \textbf{28.0}&\textbf{30.4}& \textbf{30.2} & -2.0 & -2.7 & \textbf{14.0}   &  \textbf{13.9} \\ 
	
	\hline $s_4(t)$ & 15.7& \textbf{19.4} & \textbf{22.1}& \textbf{20.4} &\textbf{23.9}& -1.8 & -3.6 & 9.3  &  11.3 \\ 
	
	\hline\hline mean & 12.5 & \textbf{23.1} & \textbf{23.5} & \textbf{25.2} & \textbf{25.5} & -1.1 & -3.8 & 11.0 & \textbf{13.4} \\
	
	\hline 
\end{tabular} 


\end{table*}

\subsection{Wideband Gaussian sources}\label{sec.con_wideband}
In many practical situations, array receives none multitone wideband signals. These signals cab be modeled as band-limited stochastic processes. In this section, we apply DDFR to these sources. 
As mentioned before in Section~\ref{sec.WDOA_estimation}, unlike the multitone scenario, $\mathbf{G}_\delta$ dictionary cannot be used here. Two group sparse solvers are used in this section, group-lasso \cite{Boyd2010a} and group gradient pursuit algorithm (group-gp) \cite{Blumensath2008}. Group-lasso deals with $\ell_{2,1}$-problem, rewritten in the Lasso form, and group-gp is a greedy algorithm to solve $\ell_{2,0}$-problem. 
\color{black}
To examine the performance in the group sparsity scenario, the proposed algorithm is compared with other wideband DOA estimators; namely $\ell_1$-SVD \cite{malioutov2005sparse}, 
W-CMSR \cite{liu2011direction}, W-LASSO \cite{Hu2012a}, 
coherent signal subspace method (CSSM) \cite{Wang1985}, incoherent MUSIC \cite{Kailatha1984}, test of orthogonality of projected subspaces (TOPS) \cite{Yoon2006a} and, Squared-TOPS \cite{Okane2010}. Note that, in CSSM a priori focusing angles are supplied through an incoherent MVDR and rotational subspace focusing matrix \cite{Kaveh1988} is used.

\begin{figure}
	\centering
	\includegraphics[width=3.0in]{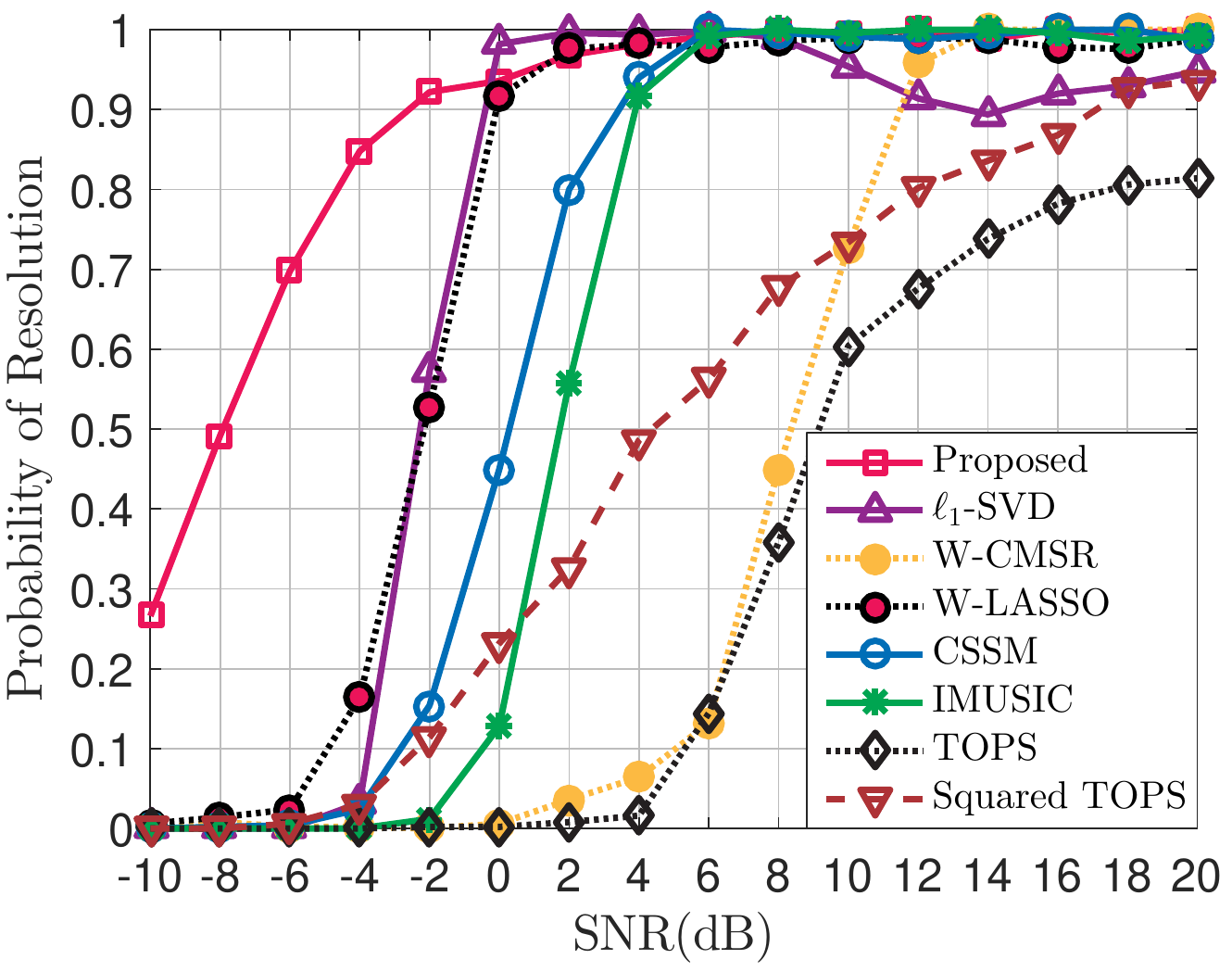}
	\caption{Separation probability versus SNR. 500 runs is used to calculate each probability.}
	\label{Fig.Separation_Prob_vs_SNR}
\end{figure}

The probability of separation is one of the common parameters to compare different methods. Two Gaussian distributed sources at $-5^\circ$ and $+5^\circ$ are considered. Sources' frequency content $f\in$[150Hz-450Hz] and a ULA with 8 omni-directional sensors and sampling frequency 2KHz is considered. Elements inter-spacing $d=\lambda_{min}/2$, where $\lambda_{min} = c/f_H$, $f_H$=450Hz and $c$=1500$m/s$. All algorithms work with 200 temporal samples and a successful result is defined as that corresponding to a spectrum with peaks that their error is less than $\pm2.5^\circ$. The probability of separation versus SNR is illustrated in Fig.~\ref{Fig.Separation_Prob_vs_SNR}. The \textit{proposed} label in Fig.~\ref{Fig.Separation_Prob_vs_SNR} refers to group-lasso applied to $\ell_{2,1}-$problem and DOA spectrum is obtained through norm-$\ell_2$ at each group. $\mathbf{G}_d$ is constructed with $\Delta\theta=1^\circ$ and $\Delta f=10$Hz. 
According to Fig.~\ref{Fig.GroupForm}, there are 181 groups (i.e. $\{-90^\circ,-89^\circ,\cdots,+90^\circ\}$) and 31 group's members (i.e. $\{100,110,\cdots,400\}$). 
The proposed solution shows superior performance at low SNR regime.
A slight drop in performance is seen for $\ell_1$-SVD at SNR$\ge$10dB, which is due to SNR-dependent behavior of the noise boundary in the $\ell_1$-SVD method (see page 7 and section VIII.B in \cite{malioutov2005sparse} for more details).

\begin{figure*}[!t]
	\centering
	\subfloat[]
	{\includegraphics[width=0.45\linewidth]{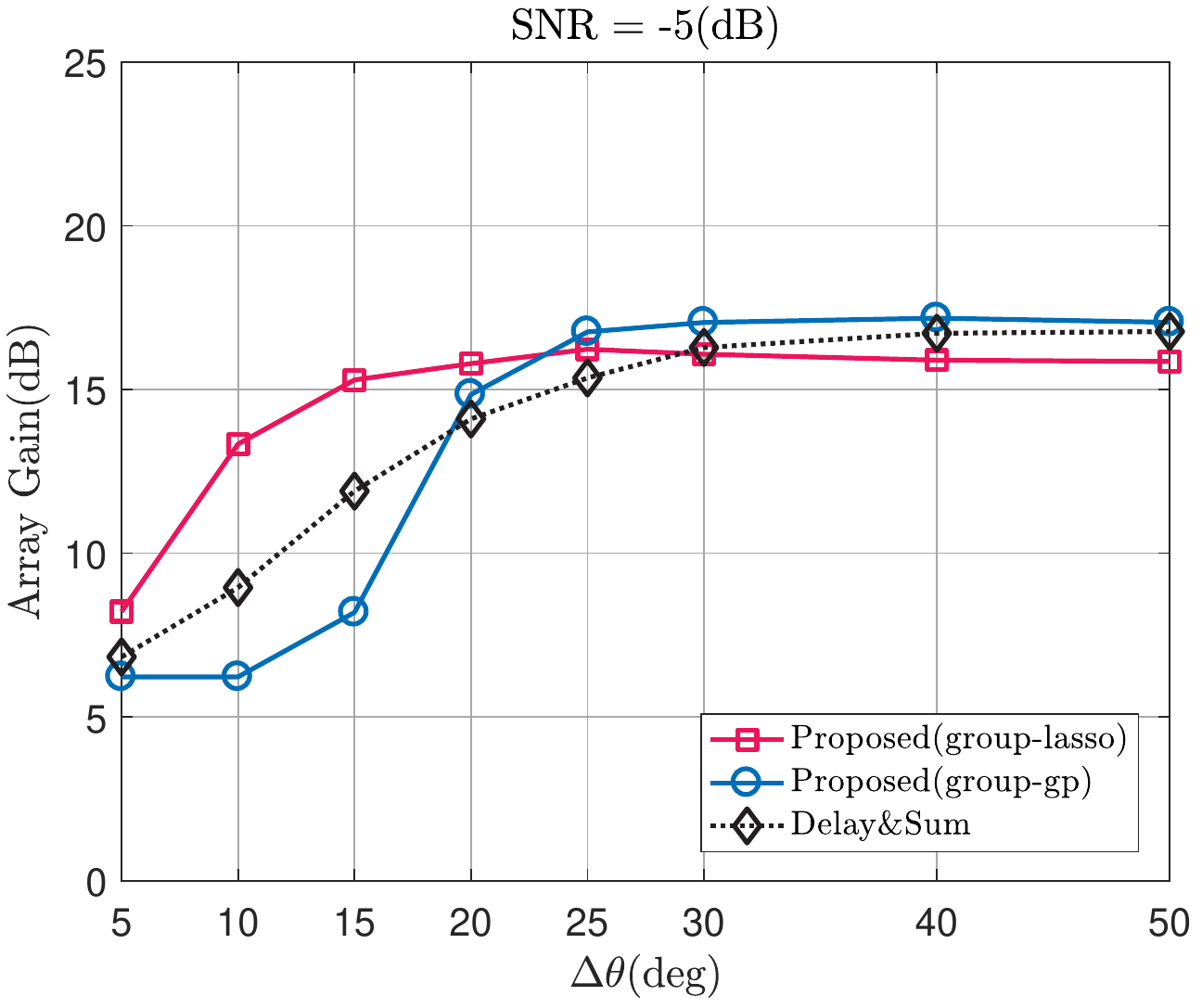}
		\label{subfig.SNR-5}}
	\hfil
	\subfloat[]
	{\includegraphics[width=0.45\linewidth]{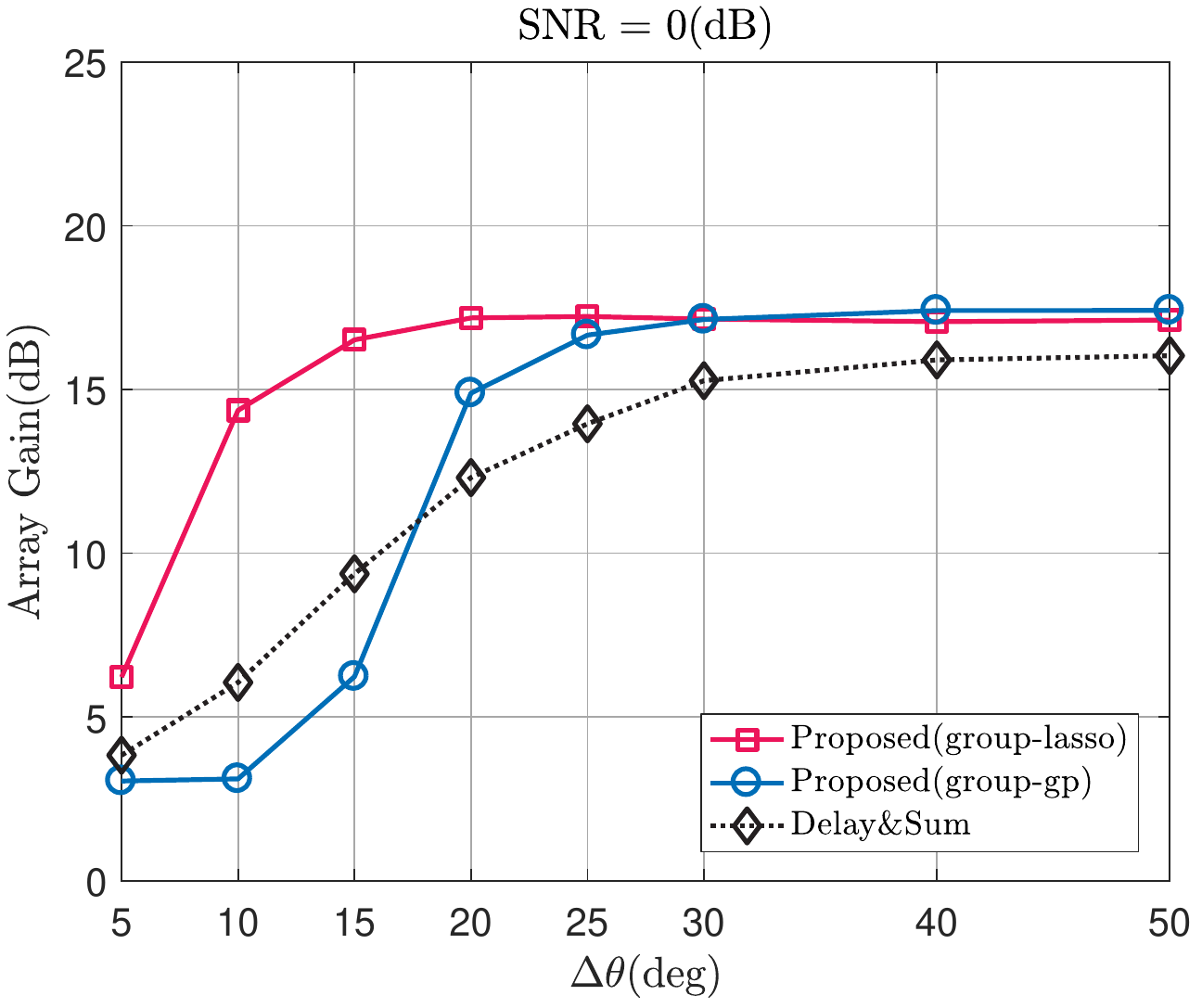}
		\label{subfig.SNR0}}
	\\
	\subfloat[]
	{\includegraphics[width=0.45\linewidth]{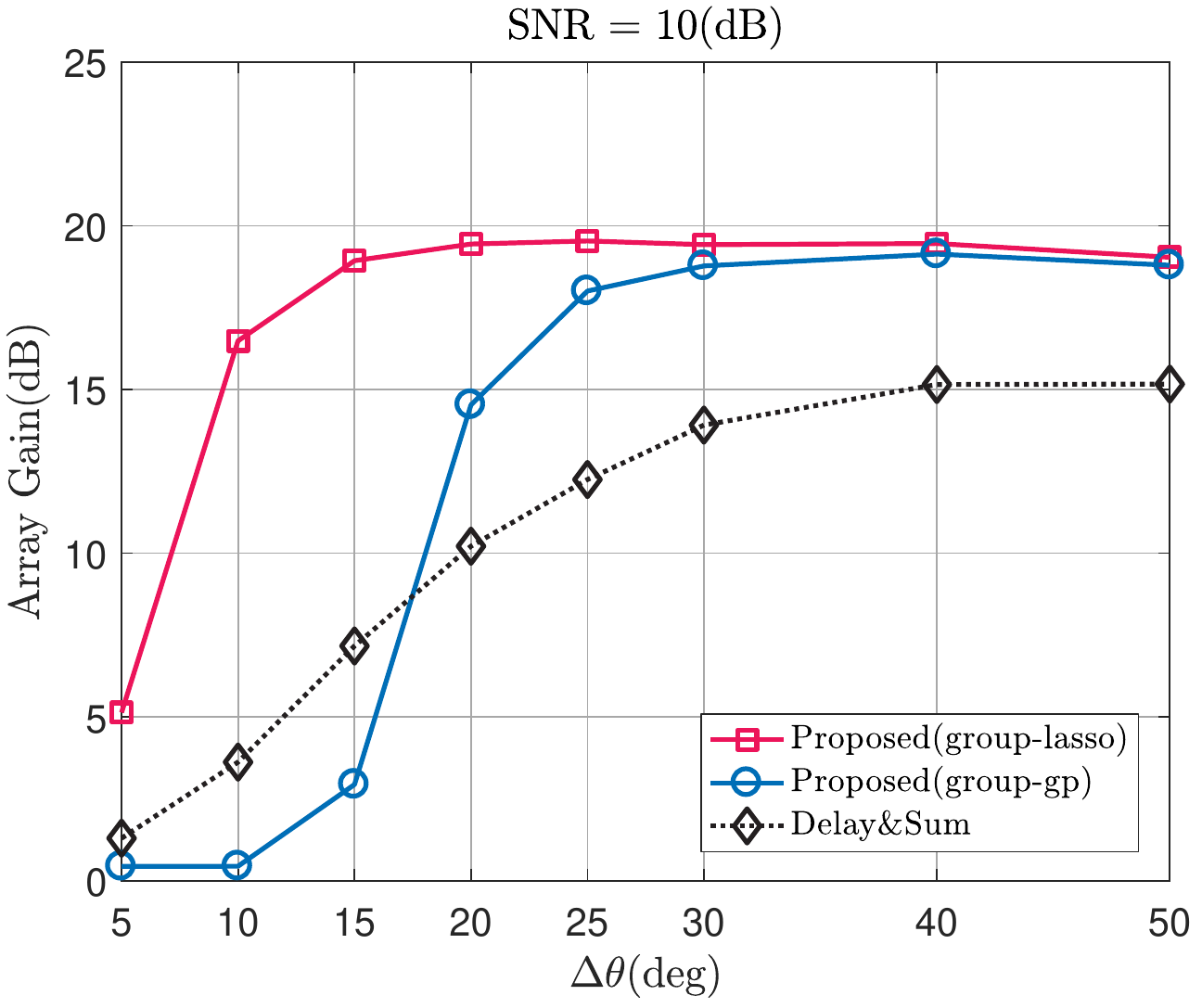}
		\label{subfig.SNR10}}
	\hfil
	\subfloat[]
	{\includegraphics[width=0.45\linewidth]{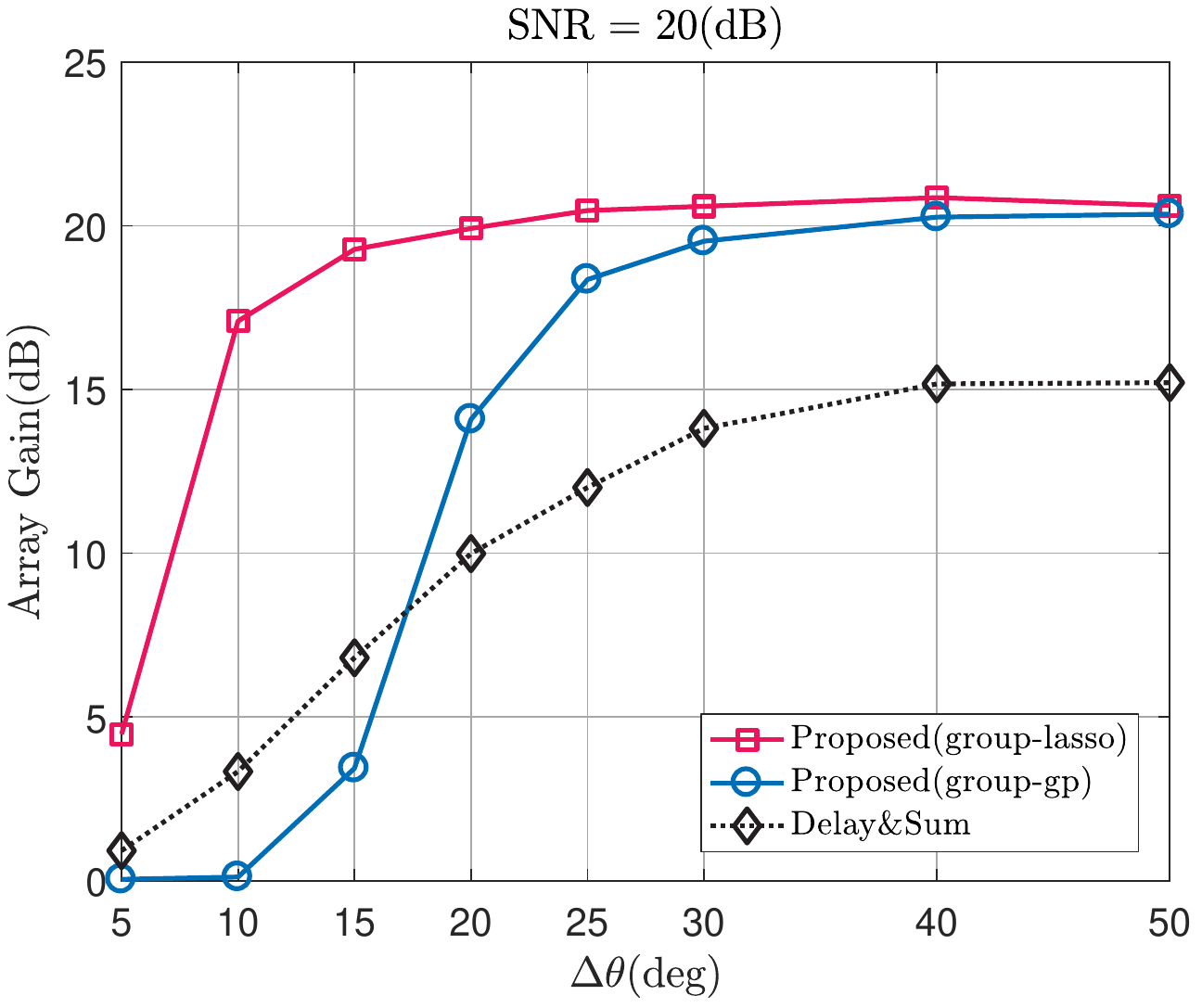}
		\label{subfig.SNR20}}
	
	\caption{Array gain versus sources $\Delta\theta$ for four SNR values \protect\subref{subfig.SNR-5}-5dB, \protect\subref{subfig.SNR0}0dB, \protect\subref{subfig.SNR10}10dB and, \protect\subref{subfig.SNR20}20dB. 100 runs are used for simulation.}
	\label{Fig.ArrayGain_vs_DeltaDOA}
\end{figure*}

In the next simulation, array gain versus sources spatial distance is examined. 
The purpose of this example is to highlight the unique capability of source recovery simultaneous with DOA estimation of the proposed method in comparison with other methods.
Simulation parameters are the same as before, but one source is fixed at $0^\circ$ and the second is placed at $\Delta\theta$. Array gain for delay\&sum, group-gp and group-lasso are calculated at each run. The simulation results are shown in Fig.~\ref{Fig.ArrayGain_vs_DeltaDOA}. At low SNR scenario, the array gain is dominated by omni-directional noise rejection ratio. Fig.~\ref{subfig.SNR-5} shows that group-lasso has superior performance for closely located sources and all three methods tend to show identical array gain for well separated sources. 

As the sources' SNR increases, array gain enhancement is dominated by interference rejection ratio. In this case, the simultaneous beamforming capability of the proposed DOA estimator has superior performance. Moreover, group-lasso is outperforming group-gp solver, although at the expense of more computational complexity.

Finally, a simulation is done to compare the computational load of the proposed method. 
The simulation parameters are similar to the second example (probability of resolution versus SNR shown in Fig.~\ref{Fig.Separation_Prob_vs_SNR}). The number of snapshots $M$ is set to 100 and 400 and the runtime is calculated for computation time of spatial spectrum in -$90^\circ$ to +$90^\circ$ with grid size $1^\circ$. This simulation is run on a PC with specifications: Windows 10 (64bit version), Intel
Core i7-920 3.4 GHz and 8GB RAM. Results are listed in Table~\ref{table.Runtime_vs_M}.

As previously mentioned, the runtime in the DDFR framework is heavily dependent on the selection of the numerical solver, but as shown in Table~\ref{table.Runtime_vs_M} a general incremental trend of the runtime is observed for both solvers (namely group-lasso and group-gp) with increasing the number of snapshots. 
In brief, the proposed framework computational cost rapidly increases with the number of snapshots as well as grid points and number of sensors. 
\begin{table*}[!t]
	\scriptsize
	\centering
	\begin{threeparttable}[b]
		\caption{Comparison of the normalized averaged runtime for different number of snapshots.}
		\label{table.Runtime_vs_M}
		\begin{tabular}{>{\centering}m{1.5cm} 
				>{\centering}m{2em} 
				>{\centering}m{2em} 
				>{\centering}m{2em} 
				>{\centering}m{2.5em} 
				>{\centering}m{2em} 
				>{\centering}m{2em} 
				>{\centering}m{2em} 
				>{\centering}m{2.5em} 
				>{\centering}m{2.5em}  
				} 
\toprule[1.0pt]
Number of snapshots  & group-lasso & group-gp & CSSM & IMUSIC & TOPS & Squared TOPS & $\ell_1$-SVD & W-CSMR & W-LASSO
  \tabularnewline
\midrule[1.0pt]
M=100\tnote{$\dag$} & 131 & 6 & 4 & 1 & 5 & 5 & 333 & 64 & 47
\tabularnewline
\midrule
M=400\tnote{$\dag$} & 306 & 21 & 4 & 1 & 5 & 5 & 305 & 63 & 41
\tabularnewline
\bottomrule[1.0pt]
\end{tabular} 

\begin{tablenotes}
\item[$\dag$] All values are normalized to the minimum runtime 11.1(msec) for M=100 and 11.6(msec) for M=400, which corresponds to the IMUSIC method.
\end{tablenotes}
	\end{threeparttable}
\end{table*}
\section{Conclusion}\label{sec.Conclusion}
A mathematical framework for representing the array temporal samples in the DOA-frequency domain, named DDFR, is proposed. This representation is formulated as a linear system, in which the concatenated observation vector lies in column space of a $\mathbf{G}$ dictionary.
 
Then, two approaches for construction of  $\mathbf{G}$ were proposed; constant $\Delta\delta$ and direct synthesize schemes. 
It is shown that $\mathbf{G}$ dictionary generation with constant $\Delta\delta$ method, enjoys the property of constant mutual coherence in iso-frequency atoms. In this regard, a dictionary design procedure was proposed that ensures a given mutual coherence. On the other hand, $\mathbf{G}$ with constant $\Delta\delta$ method has no control over DOA grid points, which may be undesirable in some applications.
DDFR does not require any subband processing or covariance estimation since it directly processes array time samples. Furthermore, it enjoys the unique property of source reconstruction simultaneous with DOA-frequency representation.
\\
Furthermore, it was shown that DDFR with imposing group sparsity constraint results in a wideband DOA estimator for non-multitone signals. This also enjoys the exclusive feature of simultaneous beamforming with DOA estimation, since it provides sources' reconstruction coefficients without any additional computations. Accordingly, DDFR is an appealing framework for further studies on simultaneous wideband DOA estimation and beamforming solutions.

\vspace*{-10pt}
\section*{References}

\end{document}